\documentclass[aps,prb,twocolumn,superscriptaddress,showpacs]{revtex4}

\usepackage{amsmath}
\usepackage{graphicx}
\usepackage[usenames,dvipsnames]{color}

\begin{document}

% ******************************
% Title of paper
% ******************************
\title{ Benchmark of Exchange-Correlation Functionals for High Pressure Hydrogen using Quantum Monte Carlo}

% ******************************
% Authors / affiliations
% ******************************
\author{Raymond C. Clay}
\email[]{rcclay2@illinois.edu}
\affiliation{Department of Physics, University of Illinois at Urbana-Champaign, Urbana, Illinois 61801, USA}

\author{Jeremy Mcminis}
\affiliation{Lawrence Livermore National Laboratory, Livermore, California 94550, USA}

\author{Jeffrey M. McMahon}
\affiliation{Department of Physics, University of Illinois at Urbana-Champaign, Urbana, Illinois 61801, USA}

\author{Carlo Pierleoni}
\affiliation{Department of Physical and Chemical Sciences, University of L'Aquila and CNISM UdR L'Aquila, Via Vetoio, I-67010 L'Aquila, Italy}

\author{David M. Ceperley}
\affiliation{Department of Physics, University of Illinois at Urbana-Champaign, Urbana, Illinois 61801, USA} 

\author{Miguel A. Morales}
\affiliation{Lawrence Livermore National Laboratory, Livermore, California 94550, USA}

% ******************************
% Date
% ******************************
\date{\today}

% ******************************
% Abstract
% ******************************
\begin{abstract}
 The ab-initio phase diagram of dense hydrogen is very sensitive to errors in the treatment of electronic correlation.  Recently, it has been shown that the choice of the density functional  has a large effect on the predicted location of both the liquid-liquid phase transition and the solid insulator-to-metal transition in dense hydrogen.  To identify the most accurate functional for dense hydrogen applications, we systematically benchmark some of the most commonly used functionals using Quantum Monte Carlo.  By considering several measures of functional accuracy, we conclude that the van der Waals and hybrid functionals significantly out perform LDA and PBE.  We support these conclusions by analyzing the impact of functional choice on structural optimization in the molecular solid, and on the location of the liquid-liquid phase transition. 

\end{abstract}


%\keywords{}

\maketitle

%%%%%%%%%%%%%%%%%%%%%%%%%%%%%%%%%%%%%%%%%%%%%%%%%%
\section{Introduction}
%%%%%%%%%%%%%%%%%%%%%%%%%%%%%%%%%%%%%%%%%%%%%%%%%%

As a consequence of being the most abundant element in the universe, hydrogen can be readily observed to exist over a very large range of temperatures and pressures.  Additionally, for many applications, such as inertial confinement fusion or modeling Jovian planets, the pressures and temperatures of hydrogen can vary by orders of magnitude within a given system.  Understanding the behavior of such systems necessitates an accurate phase diagram over large range of thermodynamic conditions, which by some estimates needs to be accurate to at least 1\% \cite{McMahon2012}.

Attaining this level of accuracy is an ongoing challenge.  Despite hydrogen's simplicity, significant nuclear quantum effects (NQE's) and electronic correlation cooperate to give rise to a rich phase diagram.  There are known to be four insulating molecular solid phases, an insulating molecular liquid phase, and a conducting atomic liquid phase at high temperatures \cite{Weir1996}.  Most interestingly, theory predicts the existence of a low-temperature metallic phase above 350 GPa, which could be a superconductor with an unusually high $T_c$ or even something more exotic \cite{Babaev2004, Moulopoulos1991, Bonev2004a, McMahonSC2012}.  

Though a rough qualitative agreement between experiment and theory has been reached for most of the hydrogen phase diagram, \textit{ab-initio} methods have struggled quantitatively in the high-pressure regime of hydrogen.  This is because the relevant energy scales in high-pressure hydrogen are comparable to the errors introduced by approximating NQE's and electronic correlation.  Recent simulations show that using the quasiharmonic approximation (QHA) produces noticeable quantitative differences from an exact treatment in quantities like pair correlation functions and enthalpies of solid hydrogen structures \cite{Morales2013a}.  Additionally, the approximation of electronic correlation through the use of density functionals has been shown to have a large effect on the predicted phase diagram.  For example, the location of the liquid-liquid phase transition (LLPT) in hydrogen could change by as much as 160 GPa in path integral molecular dynamics (PIMD) simulations depending on whether PBE or vdW-DF2 was used \cite{Morales2013}.  In the solid phase, it was found that by using either optB88-vdW or vdW-DF2 in PIMD, noticeably better agreement with experiment was obtained for the band gap in the C2/c phase and the transition pressure between phases II and III as compared to PBE \cite{Morales2013a, Li2013}.

Recent algorithmic and computational advances allow a systematic treatment of NQE's, for example with PIMD or PI+GLE \cite{Ceriotti2011}.  However, there have been, to date, few systematic studies to identify the most accurate functional for high-pressure hydrogen \cite{Azadi2013a,Azadi2013b}.  PBE has been advocated on error-cancellation grounds, but several previous studies indicate that some functionals are more accurate than the semi-local functionals.  As dispersion forces are significant in solid hydrogen, it is expected that non-local functionals that include this effect will calculate the total and relative energies more accurately, hence previous interest in vdW-DF2 and optB88-vdW. Additionally, hybrid functionals such as HSE are known to give better band-gaps than semi-local functionals, hence the interest in these functionals to describe the insulator to metal transitions in high-pressure hydrogen \cite{Morales2013}. Because of technological limitations, there is little high-quality experimental data available for pressures higher than 200 GPa, so identifying accurate functionals must be done using theoretical methods.

The purpose of this work is to systematically benchmark several types of semi-local, non-local van der Waals, and hybrid functionals against fixed-node projector quantum Monte Carlo (QMC).  QMC is an accurate many-body method that has proven to be exceptionally accurate in the study of hydrogen, making it suitable as a reference in regions where experimental data is sparse.  In section II and in the supplementary materials, we provide details of the methods used.  In section III, we quantitatively establish that non-local density functionals have superior energetics but poorer pressure estimation as compared with LDA and PBE, justifying the results of Ref. \onlinecite{Li2013} and Ref. \onlinecite{Morales2013a}. We then compare how functional choice affects the predicted hydrogen structures by looking at bond lengths, QMC enthalpies, the shape of the intramolecular potential and the location of the liquid-liquid phase transition.  In section IV, we draw our conclusions.  We have also included supplementary information, covering computational details, and providing  detailed tables of density functional performance.  

%%%%%%%%%%%%%%%%%%%%%%%%%%%%%%%%%%%%%%%%%%%%%%%%%%
\section{Method} %Computational Details}
Gauging the accuracy of density functionals has been typically done based on experimental data, for example the G2 test set of molecular binding energies \cite{Curtiss1997}.  In contrast, our method is fully \textit{ab initio}, and tailored to dense hydrogen. We used DFT-PIMD simulations to generate various ``test sets'' for a variety of phases and thermodynamic conditions as described below.  After establishing reference energies and pressures with QMC, we performed DFT calculations on our test set using up to 10 different density functionals.  Various measures were used to quantify functional accuracy.  Full  details are given in the supplementary information. First we outline the quantum Monte Carlo methods and then the various test sets and error estimates.

\label{sec:comp_details}
%%%%%%%%%%%%%%%%%%%%%%%%%%%%%%%%%%%%%%%%%%%%%%%%%%

\subsection{Quantum Monte Carlo calculations}

Several different QMC techniques were used.  
The quantum Monte Carlo Package (QMCPACK)  \cite{Esler2012a, Kim2012} was used in all diffusion Monte Carlo (DMC) and variational Monte Carlo (VMC) simulations. The Born-Oppenheimer Path Integral Monte Carlo (BOPIMC) code \cite{LNP2006,Morales2010,MoralesPNAS2010,Liberatore2011} was used in all reptation quantum Monte Carlo (RQMC) simulations. For all solid and liquid structures considered in this work, we first performed a wave function optimization using VMC. We used Slater-Jastrow type wavefunctions with backflow transformation in all cases. The single particle orbitals were obtained from a PBE-DFT calculation using the Quantum Espresso \cite{Giannozzi2009} package. This was followed by DMC calculations, within the fixed-node approximation. For a subset of the solid configurations we use RQMC, combined with correlated sampling, to calculate energy differences as we varied the bond length of the molecules in the solid.

We used the virial estimator to calculate pressures directly from QMC, which for the Coulomb interaction has the form: $P=\frac{1}{3\Omega}(2T+V)$, where $T$ ($V$) is the kinetic (potential) energy and $\Omega$ is the simulation (supercell)  volume. In order to minimize systematic errors in the calculation, we rewrite the estimator as $P=\frac{1}{3\Omega}(E+T)$, where $E$ is the ground state energy, and use an extrapolated estimator for the kinetic energy\cite{CeperleyDMandKalos1979}, $T_{extrap} = 2T_{DMC} - T_{VMC}$, which produces a pressure that is correct to second-order in the quality of the trial wavefunction.

Controllable errors, such as time-step error, projection time, and population bias, were reduced to be comparable to our desired statistical error, giving an accuracy of approximately 0.01mHa/proton for energy and 0.3GPa for the pressure estimates.  Finite-size effects were handled through a combined use of twist-averaged boundary conditions and post-processing corrections, which are detailed in the supplementary information.  The only source of uncontrollable errors come through the use of the fixed-node approximation, which we expect to be very small for hydrogen.  %We estimate this error to be approximately ???mHa/proton based on our choice of trial wavefunction.

\subsection{Test Sets }

We define a  test set $S$ as  a set of $M$ proton configurations, $R_{S}=\lbrace{\mathbf{R}_1,...\mathbf{R}_M}\rbrace$  all at a given density, temperature  and phase (liquid/solid).  To broadly classify a functional's accuracy, it is important that these test sets consist of uncorrelated but physically reasonable configurations, representative of the state of hydrogen at high pressures. For this purpose, we use constant volume PIMD simulations based on DFT. In particular, we use the PI+GLE algorithm of M. Ceriotti, \emph{et al.} \cite{Ceriotti2011}, which has been shown to significantly accelerate the rate of convergence of PIMD calculations with respect to the number of imaginary time slices.  By using PIMD to generate our configurations we ensure that we adequately sample both nuclear quantum effects and thermal fluctuations, which are critical for a proper description of hydrogen. Note that for hydrogen at 1000K,  NQE are larger than thermal effects.  Care was taken to ensure that the selected configurations were statistically independent and well equilibrated.
    
For the molecular solid, we considered the following structures obtained in previous structure searching studies:  C2/c, Cmca-12, Pbcn \cite{PickardNeeds2007}, and Cmca \cite{Edwards1996}.  For each of these structures, we first obtained zero temperature reference configurations from DFT, at pressures of $P^{DF}$=200 GPa and $P^{DF}$=300 GPa, by performing structural relaxation using the vdW-DF2 functional. The band gaps calculated with this functional were in good agreement with experimental measurements when nuclear quantum effects are taken into account \cite{Morales2013b}. Starting from these configurations, PIMD simulations were performed at a temperature of T=200 K, also with the vdW-DF2 functional. All the simulations in the solid, and hence the configurations in the resulting test sets, were performed with 96 protons. PIMD+GLE simulations in the solid were performed with a modified version of the Vienna Ab-Initio Simulation Package (VASP) \cite{Kresse1993, Kresse1994,Kresse1996,Blochl1994}. We used the projector augmented wave representation of VASP and a 3x3x3 k-point grid in the simulations. For each combination of pressure and structure, we generated configuration sets with at least 20 configurations. It is important to realize that, while we refer to the configuration sets by a structure and a pressure, these are really constant density sets. In addition, the density corresponds to that of the static lattice optimized at the given electronic pressure; thermal and quantum effects are not included during the structure optimization. While the difference in density for different structures optimized at a given pressure is small, it is important to realize that $P^{DF}$ is simply a convenient label unless otherwise indicated.  

For liquid hydrogen, we considered configurations of 54 protons at three densities given by $r_s=$ 1.30, 1.45, 1.60, where $r_s$ is the Wigner-Seitz radius, and a temperature of T=1000 K. The three densities correspond to a fully dissociated atomic liquid, a molecular liquid in the neighborhood of the liquid-liquid phase transition and a fully molecular liquid, respectively. These test sets allow us to compare the performance of DFT functionals in different environments in the liquid. The PIMD+GLE simulations in this case were performed with a modified version of QE and the vdW-DF2 functional. We used a Troullier-Martins norm-conserving pseudo-potential with a cutoff of 0.5 $Bohr$ and a 2x2x2 k-point grid. Test sets in the liquid were generated with approximately 100 configurations at each density.

Finally, we used an enlarged set of zero-temperature configurations in order to study the influence of the choice of functional on structural properties in the solid.  Specifically, we study the following structures from previous structure searching studies:  C2/c, Cmca-12, Pbcn \cite{PickardNeeds2007}, Cmca \cite{Edwards1996}, and mC24-C2/c \cite{Liu2012}.  Ground state structures were relaxed using the PBE, vdW-DF, and vdW-DF2 functionals at $P^{DF}=$ 200 GPa, 300 GPa and 400 GPa.  

For additional details on the configuration sets used in this work, see the included Supplementary Information section.

\subsection{Density Functional Comparison}

For each configuration in a test sets in the solid, we calculated its energy and pressure using the following ten functionals:  LDA, PBE\cite{Perdew1981,Perdew1996}, vdW-DF \cite{Dion2004}, vdW-DF2 \cite{Lee2010}, vdW-optPBE, vdW-optB88 \cite{Klimes2010}, vdW-optB86B \cite{Klimes2011}, BLYP \cite{Grimme2006}, vdW-TS \cite{Tkatchenko2009}, and HSE  \cite{Heyd2003}.   We used a restricted set of functionals for the liquid test sets: LDA, PBE, vdW-DF, vdW-DF2, and HSE.  

For all $M$ configurations in a given test set $S$, we compute the density functional error of an observable $\mathcal{A}$ as $\delta \mathcal{A}^{DF} = \mathcal{A}^{DF}-\mathcal{A}^{QMC}$.  From this, we used two general measures of error.    
The first is the average error
which we define as:
\begin{equation}
\langle \delta \mathcal{A}^{DF} \rangle_S = \frac{1}{M} \sum_{\mathbf{R}_i \in S} \delta \mathcal{A}^{DF}(\mathbf{R}_i)
\end{equation}

We also use a more general error measurement, akin to a mean-absolute error, which we define as:
\begin{equation}
\langle | \widetilde{\delta \mathcal{A}^{DF}} |\rangle_S = \frac{1}{M} \sum_{\mathbf{R}_i \in S} | \delta \mathcal{A}^{DF}(\mathbf{R}_i) - c^{DF}| 
\end{equation}
Here, $c$ is an density functional dependent offset, chosen by minimizing  $\langle | \widetilde{\delta \mathcal{A}^{DF}} |\rangle_{S'}$ over some set $S'$ (which does not have to equal $S$).  In the Results section, we will devise two error measures for energies that differ in their choice of reference point, but the motivation and justification will be handled later.   %\DC{COMMENT OUT OF PLACE? For the solid configurations, we generated thermal configurations for multiple solid structures at the same pressure.  However, to assess the quality of a functional, we average the various deviations measures over all structures at a given pressure. }

%%%%%%%%%%%%%%%%%%%%%%%%%%%%%%%%%%%%%%%%%%%%%%%%%%
\section{Results}
\label{sec:results}
%%%%%%%%%%%%%%%%%%%%%%%%%%%%%%%%%%%%%%%%%%%%%%%%%%
\subsection{Benchmarking}
Our concern in this section will be to establish how well various functionals capture different features of the  Born-Oppenheimer (BO) potential energy surface;  i.e. how well a functional calculates the electronic energy as a function of the proton positions.   We will facilitate this discussion by distinguishing between two concepts:  ``global energetics" and ``local energetics".  Additionally, we complement this discussion by considering how well density functionals estimate pressures.  

``Global energetics" will refer to how well a given functional is capable of capturing energy differences between arbitrary configurations at a fixed density.  This is relevant in deciding which structure from a list of different structures has the lowest energy at a fixed density.  

``Local energetics" will refer to how well a functional captures energy differences between a structure and perturbations around that structure, at a fixed density.  In the case of a solid where a well-defined local minimum exists in the potential energy surface, we need to describe the shape of the potential energy well as accurately as possible.  This is relevant for applications involving thermal and quantum fluctuations, such as the calculation of phonons and vibrational spectra, since what is relevant are energy differences between closely related structures.  Thus, the ability of a functional to accurately capture these small energy differences will affect the quality of those predictions. 

The previous two definitions refer to how well different features of the BO energy surface are captured at a fixed density.  When determining phase stabilities however, it is often important to compare energies between structures at different densities.  By using basic thermodynamic identities, we can relate the error in the pressure to how much the energy error changes between similar structures at slightly different densities: 
\begin{equation}
 \frac{\partial}{\partial \rho} \delta e^{DF} = \rho^{-2} \delta P^{DF}
 \end{equation}
We will focus on the errors in the pressure in this discussion, appealing to the above relationship.
 
\subsubsection{Global Energetics}
\label{subsec:global_energetics}
To establish which functional has the best global energetic properties, we looked at the mean-absolute errors in the energy per proton.  For a particular DFT pressure, $P^{DF}$, we build a test set $S'$ by including all the configurations consistent with that pressure, including ones with different structures.  Then, for each pressure and density functional independently, we choose a reference point $c^{DF}(P^{DF})$ by minimizing $\langle  \widetilde{|\delta e^{DF}|} \rangle_{S'}$. In practice, this amounts to choosing the median of $\delta e^{DF}$ on this aggregated test set $S'$.  
%Then, using this choice for $c^{DF}(P)$, we calculated $\langle \widetilde{|\delta e^{DF}|}\rangle_S$ over a particular test set $S$ corresponding to some solid structure at the pressure $P$. Averaging this over all structures at the given pressure $P$ gives us a measure of the error made in global capturing global energetics, which we will denote $\langle \widetilde{ |\delta e^{DF}(P) | } \rangle_{global}$
Using this choice for $c^{DF}(P^{DF})$, we calculated $\langle \widetilde{|\delta e^{DF}|}\rangle_S$ over each structure and averaged this over all structures to obtain what we denote $\langle \widetilde{ |\delta e^{DF}(P^{DF}) | } \rangle_{global}$.

\begin{figure}[h]
    \includegraphics[scale=0.40]{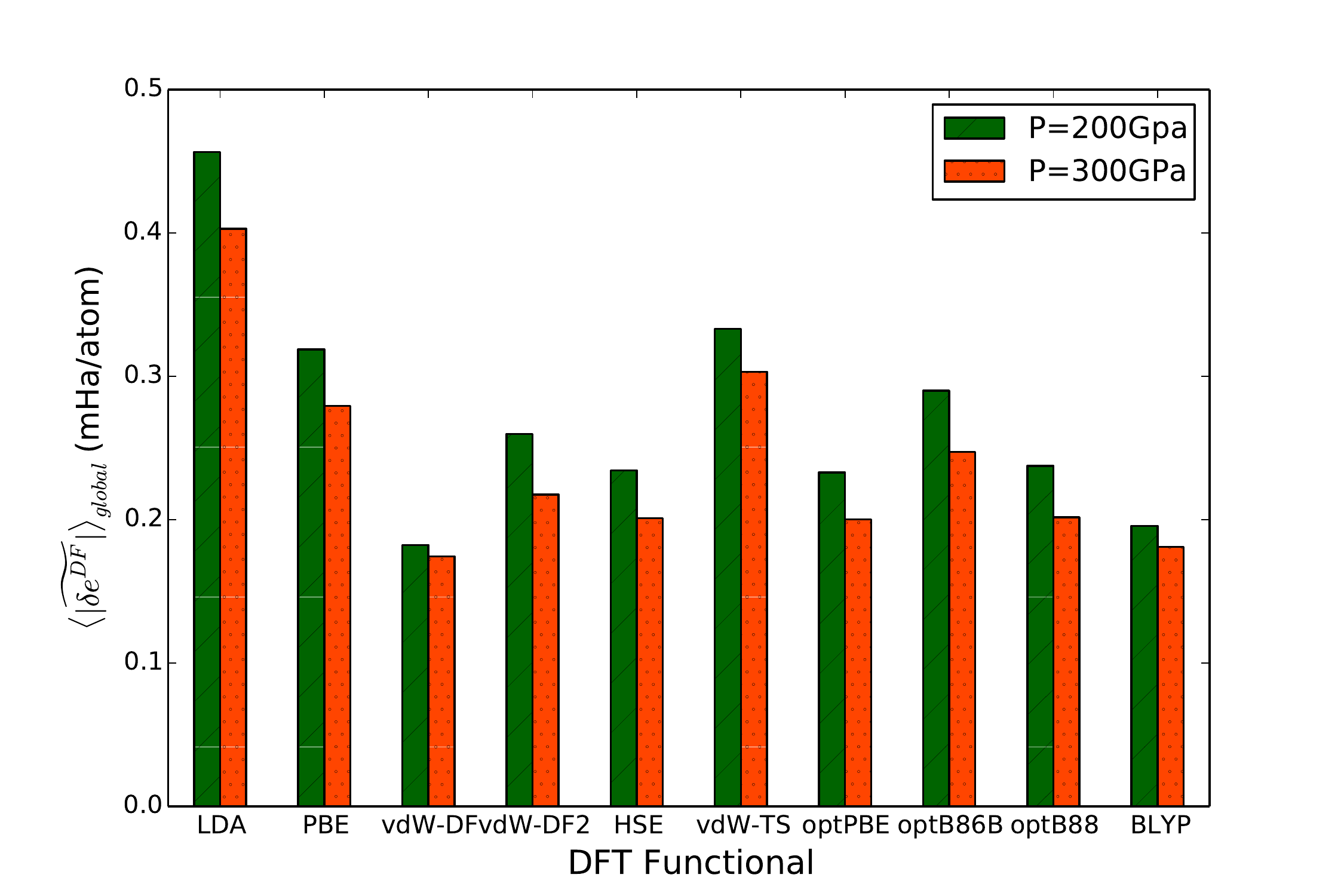}
    	\includegraphics[scale=0.40]{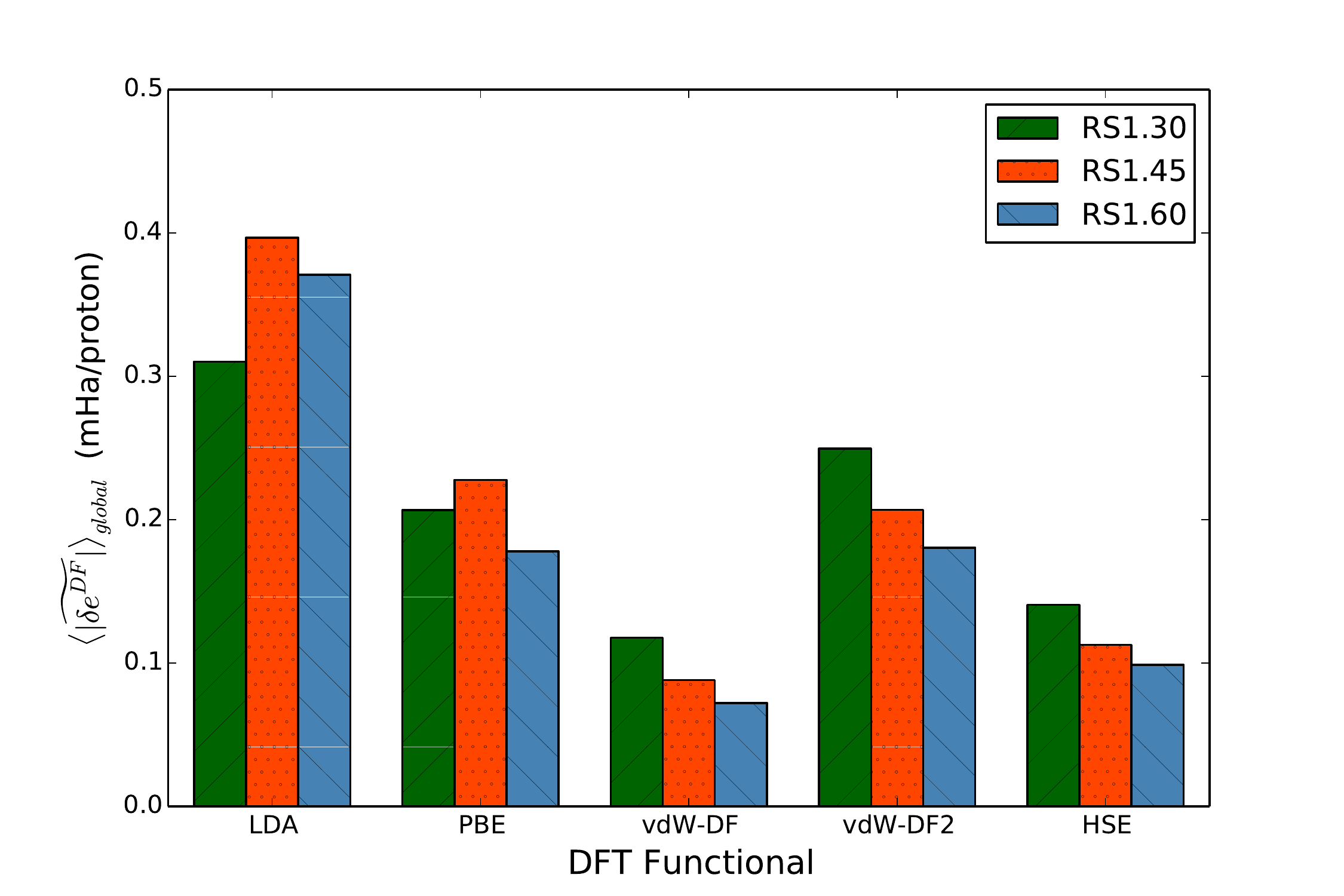}
    
     \caption{(Top) $\langle \widetilde{ |\delta e^{DF}(P) | } \rangle_{global}$ versus functional for solid molecular test sets.  Data for P=200 GPa and P=300 GPa is shown in the legend.  (Bottom)  $\langle \widetilde{ |\delta e^{DF}(P) | } \rangle_{global}$ versus functional for the liquid test sets.  Data for $r_s=$1.30, 1.45, and 1.60 shown. } 
     \label{fig:meandeviations_energy}
\end{figure}

In the top of Figure \ref{fig:meandeviations_energy}, we plot  $\langle \widetilde{ |\delta e^{DF}(P) | } \rangle_{global}$ versus functional for solid molecular hydrogen.  We included results for both the $P=200$ $GPa$ and the $P=300$ $GPa$ structures, which are marked with striped and dotted bars respectively. Two things immediately stand out.  The first is that nearly all of the hybrid and improved van der Waals functionals, excluding vdW-TS, noticeably outperform the LDA and PBE functionals.  Secondly, the vdW-DF functional seems to have the best global energetic performance out of all functionals considered, followed by BLYP and HSE. 

In the bottom of Figure \ref{fig:meandeviations_energy}, we show a plot of $\langle \widetilde{ |\delta e^{DF}(\rho) | } \rangle_{global}$ for the liquid configurations.  We have included data for the three densities $r_s$=1.30, 1.45, 1.60, which are identified in the legend.  Notice that as in the solids, vdW-DF is the best performing functional, although the hybrid functional HSE is a close runner up. 

Despite the vast differences in structures and densities, we see a very consistent picture regarding how accurate various functionals are in capturing global energetics. For the solid test set, we find that PBE is accurate to approximately 0.3 mHa/proton in dense hydrogen, whereas vdW-DF and HSE are good to 0.19 mHa/proton and 0.24 mHa/proton respectively.  The errors are smaller in the liquid phase, but the ordering of these functionals is the same for both cases with vdW-DF noticably more accurate.  
	 
\subsubsection{Local Energetics}
\label{subsec:local_energetics}

To measure the local energetics, we again used a shifted mean absolute error for the energy per proton, but with the reference point chosen to be specific to a given structure.  For a test set $S$ corresponding to a particular structure at pressure $P$, we again let the energy shift $c^{DF}$ be chosen to minimize $\langle \widetilde{ |\delta e^{DF}(P) | } \rangle_S $ on the same set S.  Averaging $\langle \widetilde{ |\delta e^{DF}(P) | } \rangle_S$ over all structures gives us a pressure dependent measure of the local energetic errors, which we will denote $\langle \widetilde{ |\delta e^{DF}(P) | } \rangle_{local}$. Notice that in this case, we are only concerned with relative errors between close configurations in the potential energy surface;  systematic shifts between the various structures are not considered.

\begin{figure}[h]
    \includegraphics[scale=0.40]{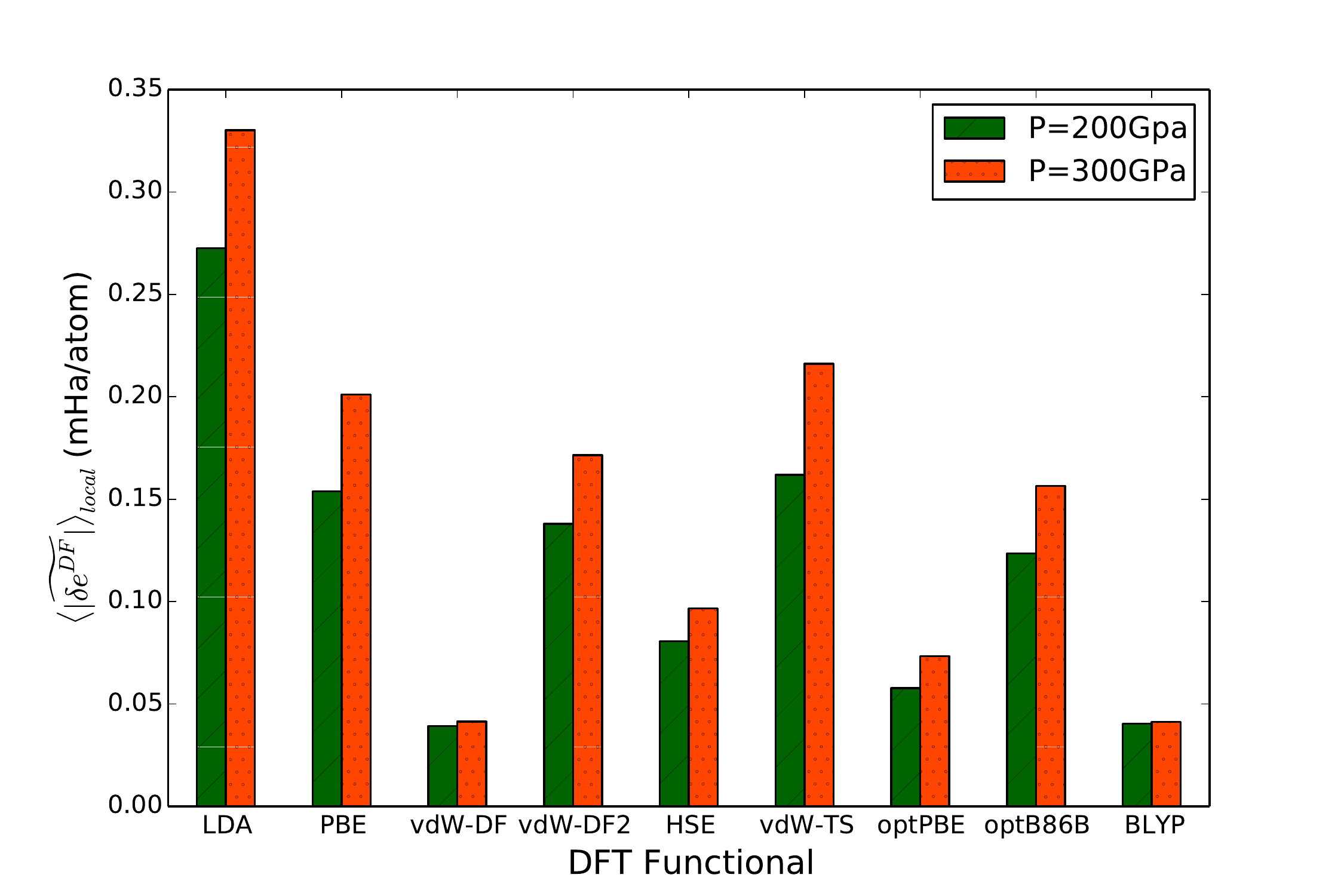}

     \caption{$\langle \widetilde{ |\delta e^{DF}(P) | } \rangle_{local}$ versus functional for solid molecular test sets.  Data for P=200 GPa and P=300 GPa is shown in the legend.  } 
     \label{fig:mae_deviations_energy}
\end{figure}

In Figure \ref{fig:mae_deviations_energy}, we show  $\langle \widetilde{ |\delta e^{DF}(P) | } \rangle_{local}$ versus density functional for solid molecular hydrogen.  The results for $P=200$ $GPa$ and $P=300$ $GPa$ are shown on the plot with dashed and dotted bars respectively. The vdW-DF functional was the most accurate  in capturing relative energy differences between similar configurations with BLYP a close second. After these functionals, the optPBE and HSE functionals exhibited fair performance.  The worst performing DF was LDA, followed by vdW-TS, and then jointly by vdW-DF2 and PBE.  This same trend was observed for the global energetic performance in Figure \ref{fig:meandeviations_energy}.

It's interesting to note how the magnitudes of the global energetic and local energetic errors compare. LDA and vdW-DF2 have local errors that are approximately 70\% the size of their global errors, and thus experience only modest accuracy gains when considering energy differences between closely related structures.   HSE and PBE perform moderately better, having local errors that are approximately 50\% and and 60\% of their global errors.  Lastly, the vdW-DF functional, beyond having the lowest magnitude of global energetic errors, experiences a local energy error that is approximately 25\% of the global errors.

In summary, we find that the van der Waals functionals were most able to calculate relative energy differences around local minima, with vdW-DF and BLYP having the smallest local energetic errors.  This might have been guessed from the previous section, as these same functionals were, on average, the best for capturing large scale energy differences. Thus, for structural relaxation, zero-point energy calculations, QMD, and other applications where location and shape of the local minimum is important, the vdW-DF functional is strongly recommended.

\subsubsection{Pressures}
\label{subsec:pressures}
\begin{figure}[h]
    \includegraphics[scale=0.40]{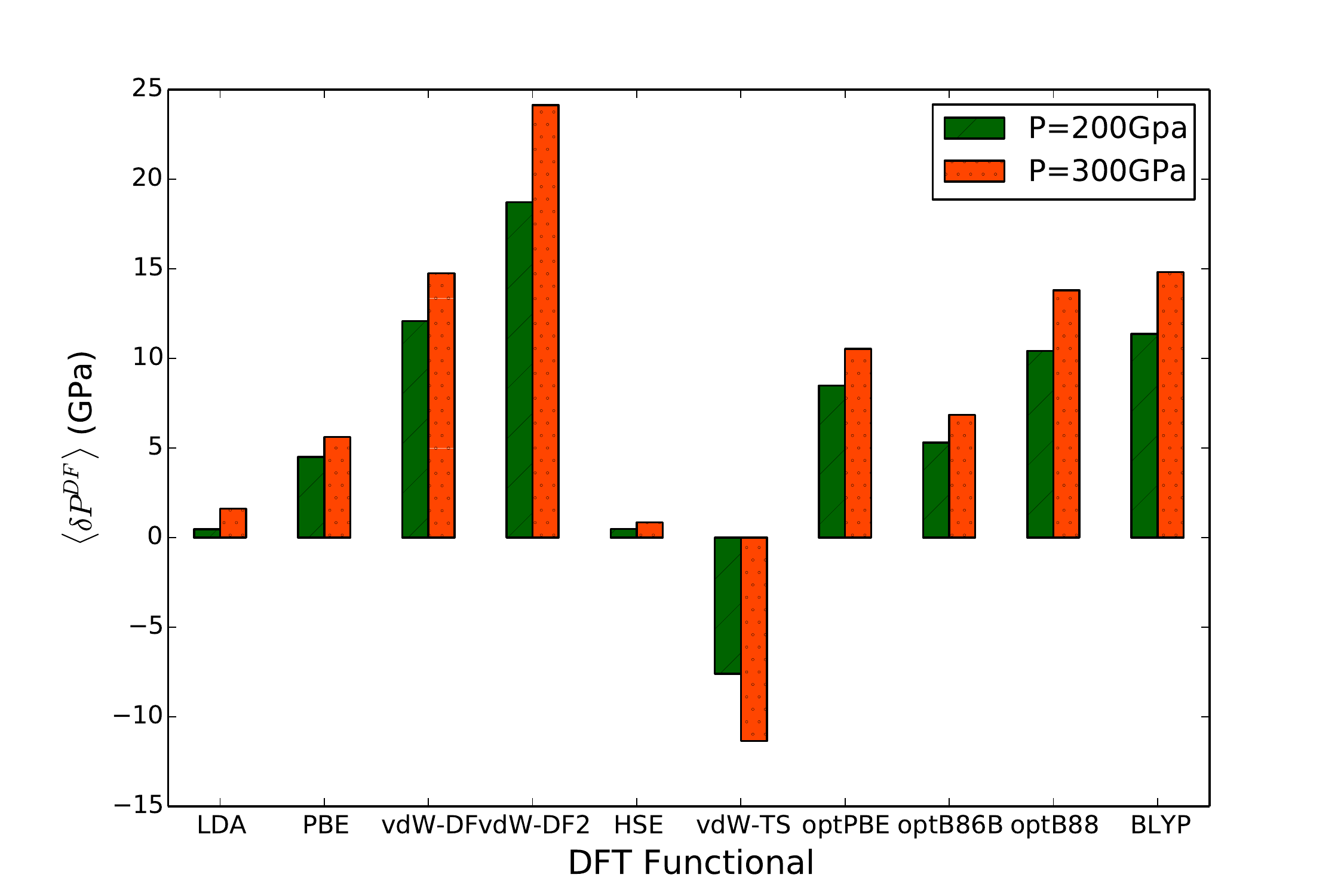}
    	\includegraphics[scale=0.40]{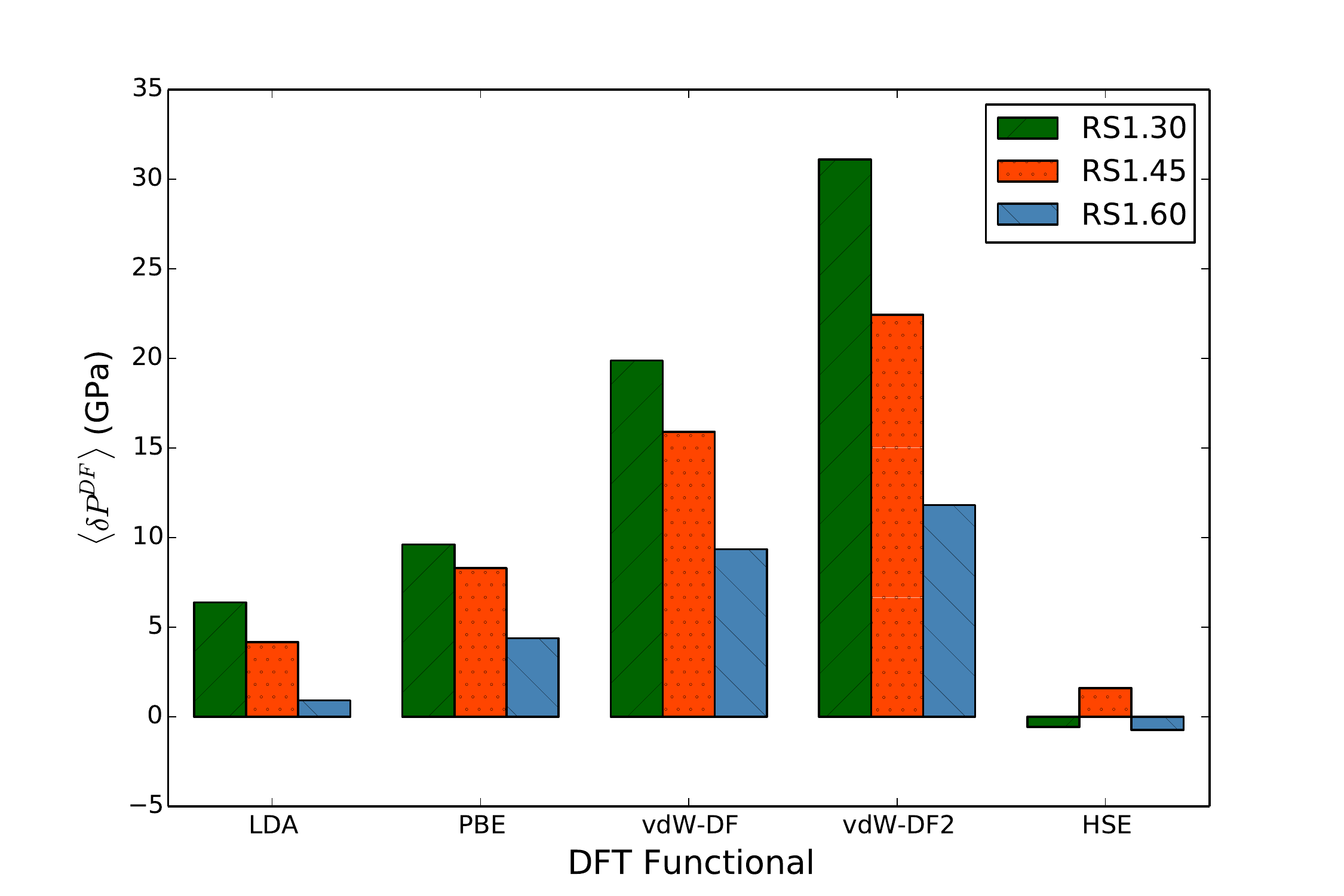}
    
     \caption{ (Top) $\langle \delta P^{DF} \rangle$ versus DFT functional for the molecular solids.  (Bottom) $\langle \delta P^{DF} \rangle$ versus DFT functional for the liquid configurations.  } 
     \label{fig:meandeviations_pressure}
\end{figure}

For a test set $S$ corresponding to a structure at particular density, we averaged $\langle \delta P^{DF} \rangle_S$  over all structures at the same pressure or density to estimate the error in the pressure. The top of Figure \ref{fig:meandeviations_pressure} shows $\langle \delta P^{DF} \rangle$ for the solids.  We see that in contrast to the local and global energetics sections, the semi-local functionals have some of the lowest pressure errors.  HSE is the best performing functional in this regard.  Note that the van der Waals functionals are among the worst performing functionals for the average pressures, with vdW-DF coming in just behind vdW-DF2 for highest pressure errors.  These observations are also seen in the bottom of Figure \ref{fig:meandeviations_pressure}, which shows $\langle P^{DF} \rangle$ for the liquid configurations.    

We also looked at $\langle | \widetilde{\delta P^{DF}}| \rangle_{local}$ for the solids and $\langle | \widetilde{\delta P^{DF}}| \rangle_{global}$ for the liquids, defined as was done with the energy errors in the local and global energetic sections.  We find that the magnitude of  $\langle | \widetilde{\delta P^{DF}}| \rangle_{local}$ across all configurations and densities is statistically indistinguishable from  the error bars of our QMC pressure estimates, indicating that the errors in the pressure are roughly independent of the configurations.  Thus, the pressure errors observed were mostly functional and density dependent constant offsets from $P^{QMC}$. Such was not the case for the energy. 

To conclude, when it comes to capturing global and local energy differences at a fixed density, including exact exchange or van der Waals effects will generally improve the energetics of density functionals for dense hydrogen.  The vdW-DF functional in particular gives noticeable improvements over PBE in capturing global energetics, and does exceptionally well for capturing local energetics.  In spite of this, HSE and the semi-local functionals outperform nearly all the van der Waals functionals when it comes to correctly estimating pressures. Given how systematic the pressure errors are, one can correct the pressure of energetically favorable DF's like vdW-DF by estimating an overall correction from either LDA or QMC. Fortunately, these errors are far more consistent than the energy errors, and so there should be some way of improving upon these functionals for future hydrogen applications.

%To conclude, for the solid molecular hydrogen in all the examined structures, the vdW-DF2, vdW-DF, and BLYP functionals most accurately capture global energy differences, excelling at higher pressures and being only slightly worse than the semilocal functionals at lower pressures.  In the liquid phase, vdW-DF or HSE are best. When it comes to pressures, the LDA, PBE, and HSE outperform all the van der Waals functionals.  This is not contradictory, as we see that the energy errors scale differently between the semilocal, hybrid, and van der Waals functionals.  Given how systematic the pressure errors are, one can correct the pressure of energetically favorable DF's like vdW-DF in post-processing, getting the magnitude of the correction from either LDA or QMC.  %UNREALISTIC OPTION? The second is to improve the functionals themselves. 

\subsection{Effects of Functional Choice}
In this section, we see how the energetic considerations of the benchmarking section relate to current problems of interest in the phase diagram of high-pressure hydrogen.  Specifically, we look at how accurately different functionals predict $H_2$ bond lengths relative to QMC optimized structures.  We also look at QMC cold curves for ground-state structures optimized with different functionals, and at the relation between the location of the LLPT and the mean absolute error of a selection of DFT functionals.

\subsubsection{Bond Lengths}
\label{subset:bond_lengths}

\begin{figure}[h]
    \includegraphics[scale=0.33]{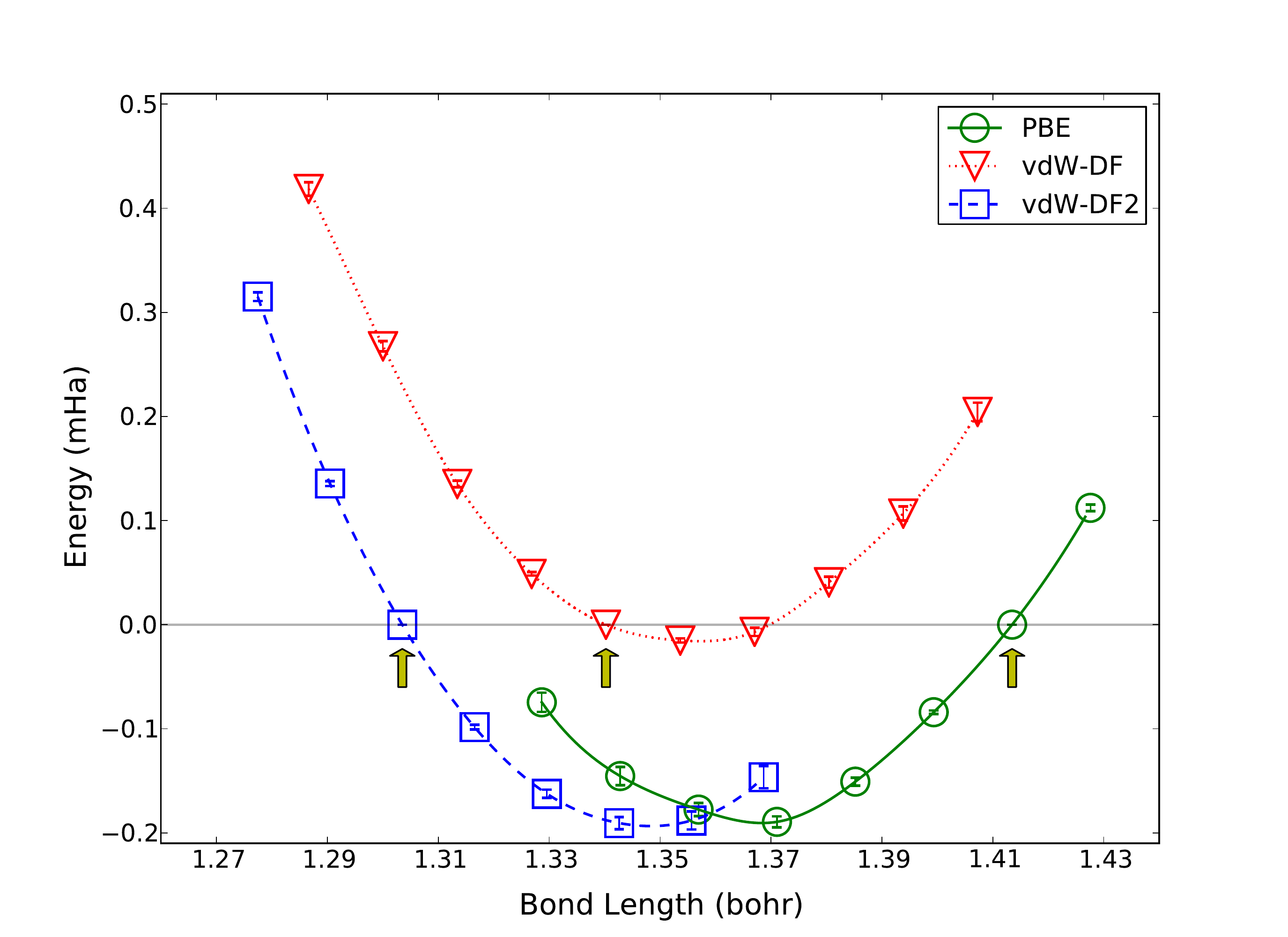}
     \caption{(Color online) Molecular bond length dependence of the QMC energy for the $C2/c$ structure at 200 $GPa$. The energy is measured with respect to the optimal bond length according to DFT.  Arrows denote the equilibrium bond length predicted by DFT.}
     \label{fig:e_vs_bl}
\end{figure}

The magnitude and pressure dependence of the bond length of the hydrogen molecule in the solid depends significantly on the DFT exchange-correlation functional. The LDA and PBE functionals predict bond lengths that are larger than that of the isolated molecule, while HSE, vdW-DF and vdW-DF2 predict bond lengths that are smaller. In order to measure the ability of each functional to predict the correct magnitude and pressure dependence of the bond length of the molecule in the solid phase, we calculated the dependence of the energy on bond length for several of the proposed solid phases of hydrogen using QMC and compared with DFT predictions. To do this we optimized a set of candidate structures (C2/c, Cmca, Cmca12, Pbcn, and mC24-C2/c) with the PBE, vdW-DF and vdW-DF2 density functionals at three different pressures: 200 GPa, 300 GPa and 400 GPa. For each combination of structure, density functional and pressure, we calculated the dependence of the energy as a function of the molecular bond length using a QMC correlated sampling technique. In particular, we calculated the change in energy produced by scaling all the molecular bond lengths in the solid by a given fraction. Figure \ref{fig:e_vs_bl} shows an example of this procedure for the C2/c structure. In this figure, the energy difference is with respect to the optimal molecular bond length according to the density functional used in the structural optimization. As described above, PBE shows a significant overestimation of the bond length, while vdW-DF2 underestimates it with a comparable magnitude. The vdW-DF functional, on the other hand, agrees well with the QMC predictions producing a structure with a negligible energy error due to the relaxation of the bond length. This relaxation energy can be significant in structures predicted by the other functionals, and, since its not guaranteed to be consistent between structures, it can significantly bias structural predictions. While the optimal bond lengths according to QMC are very similar for this structure, they depend slightly on the other structural parameters; we only relax the bond lengths in this calculation, leaving both molecular orientations and simulation cell fixed. As we discuss below this produces an additional variation in the energy of the structures, making them dependent on the functional used to optimize them, even after the bond lengths have been relaxed.

\begin{table}[t]
\begin{center}
\begin{tabular}{c c c c}
\hline
\hline
Structure &  200 ($GPa$)  &  300 ($GPa$)   & 400 ($GPa$)    \\
\hline
\hline
 \multicolumn{4}{c}{PBE [0.19]} \\
\hline
  $C2/c$  & 0.180(6) & 0.186(9) & 0.171(9) \\ 
  $Cmca$ & 0.30(1) & 0.15(2) & 0.15(1)  \\
  $Cmca-12$ & 0.185(8) & 0.24(2) & -\\
   $Pbcn$ & 0.37(1) & 0.28(2) & 0.23(2) \\
   $mC24-C2/c$ & 0.069(4) & 0.079(5) & 0.102(6) \\   
\hline  
\hline
 \multicolumn{4}{c}{vdW-DF [0.01]} \\
\hline
  $C2/c$ & 0.018(3) & 0.007(3) & 0.015(3) \\ 
  $Cmca$ & 0.011(4) & 0.007(4) & 0.017(4)\\
  $Cmca-12$ & 0.016(3) & 0.012(7) & 0.023(4) \\
   $Pbcn$ & 0.016(3) & 0.016(3) & 0.015(4) \\
   $mC24-C2/c$ & -0.005(3) & 0.001(5) & 0.008(3)  \\   
\hline
\hline
 \multicolumn{4}{c}{vdW-DF2 [0.22]} \\
\hline  
  $C2/c$ & 0.197(6) & 0.239(8) & 0.247(7) \\ 
  $Cmca$ & 0.25(1)& 0.21(1) & 0.22(1) \\
  $Cmca-12$ & - & 0.260(7) & 0.24(1) \\
   $Pbcn$ &  0.22(1) & 0.216(8) & 0.22(1) \\
   $mC24-C2/c$ & 0.222(2) & 0.18(1) & 0.22(2) \\   
\hline
\hline
\end{tabular}
\end{center}

\caption{QMC energy difference (in mHa/proton) between the structures at the optimal DFT and QMC bond lengths. Error bars are shown in parenthesis. The mean energy difference, averaged over all structures and all pressures, is shown in squared brackets next to each functional name.}
\label{Table:relax_energy}
\end{table}%

\begin{figure}[h]
    \includegraphics[scale=0.45]{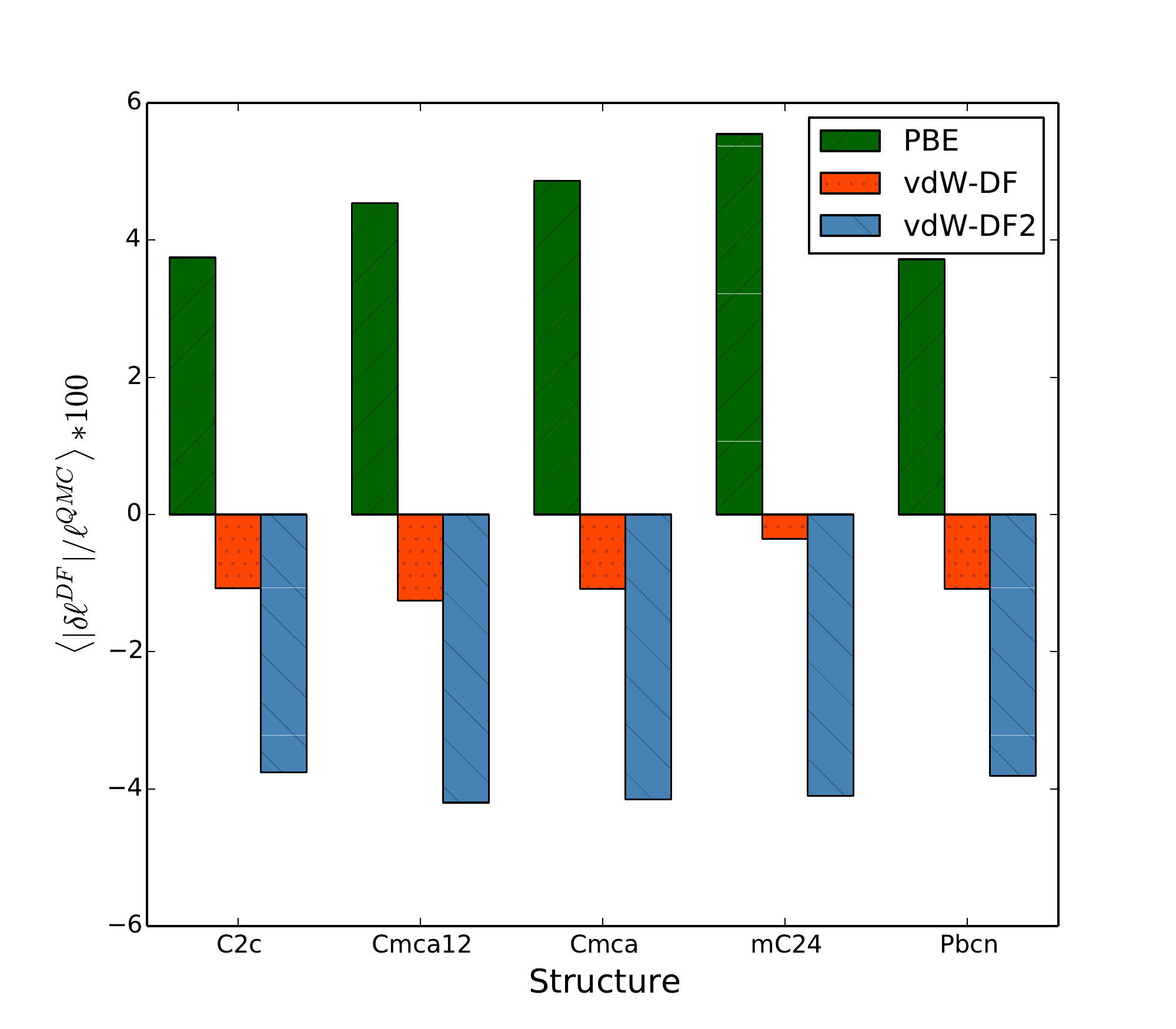}
     \caption{(Color online) Error in the molecular bond length $\ell$ of selected DFT functionals, relative to QMC optimized values. The results are averaged over all pressures considered in this work, since the pressure dependence of the error is small. }
     \label{fig:bl_vs_str}
\end{figure}

\begin{figure}[h]
    \includegraphics[scale=0.40]{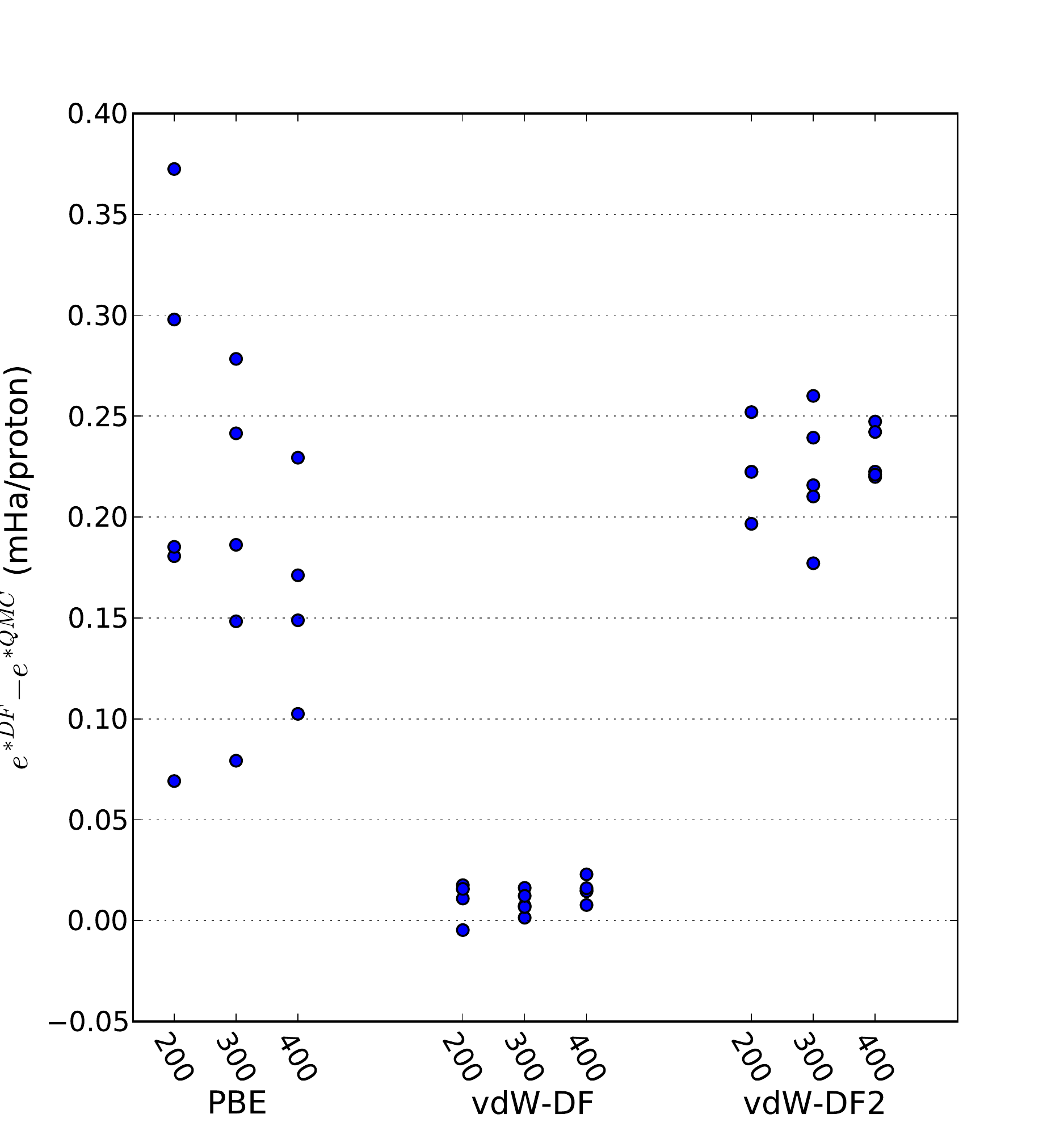}
     \caption{(Color online) Scatter plot of $e^{*DF}-e^{*QMC}$ for structures optimized with different functionals and at different $P^{DF}$.  The asterisks denote that we are measuring the QMC energy difference between a DF optimized structure and a bond-length optimized structure. The numbers on the x-axis correspond to the $P^{DF}$ in GPa at which all ground-state structures were relaxed.  The names under each triplet of numbers denote the DF used in the structural optimization.    }
     \label{fig:en_err}
\end{figure}

Figure \ref{fig:bl_vs_str}  shows the difference between the optimal QMC and DFT molecular bond length for each structure and density functional, averaged over all the pressures considered. In general, the discrepancy between QMC and DFT on the magnitude of the bond length is fairly insensitive to pressure in the range considered. Table \ref{Table:relax_energy} shows a summary of the difference in QMC energy between the optimal DFT and QMC bond lengths for all the structures, pressures and DFT functionals considered in this work.  Figure \ref{fig:en_err} shows this same data in a scatter plot, organized by $P^{DF}$ and DF.  The spread of values in the case of PBE is quite large, while the corresponding spread for vdW-DF is very small. Inaccuracies in the bond length can lead to important limitations in the predictive capabilities of DFT in this regime of the phase diagram, mainly due to the immediate proximity of this regime to a dissociation and a metal-insulator transition at higher pressures. Notice that more accurate electronic structure calculations typically require the use of structures from more approximate methods, such as DFT. Incorrect predictions of structural properties will lead to biases in the predictions of more accurate methods. On the other hand, the ordering of structures can also be significantly biased if structural parameters are not accurate, in this case the existence of molecules with large bond lengths present in many of the proposed structures for hydrogen near metallization is put into question by these calculations, since they have all been predicted using PBE which severely overestimates bond lengths.

\subsubsection{Ground State Structures}

\begin{figure}[h]
    	\includegraphics[scale=0.35]{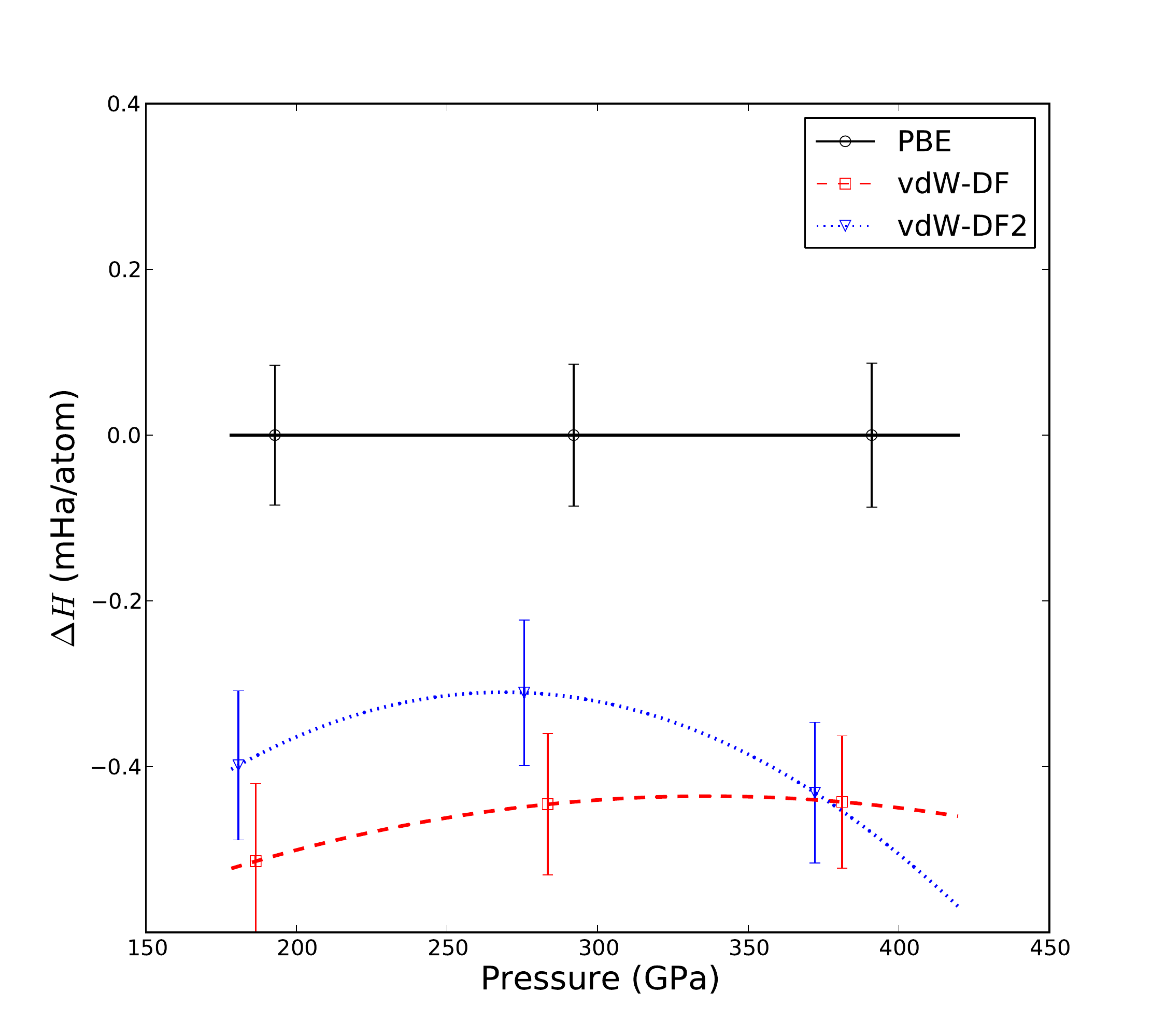} 
    
     \caption{(Color online) Relative enthalpies of $C2/c$ structures relaxed with either PBE, vdW-DF, or vdW-DF2.} 
     \label{fig:C2centh}
\end{figure}

In Figure \ref{fig:C2centh}, we show the QMC enthalpies (relative to the enthalpies of the PBE optimized structures) of $C2/c$ structures relaxed with the PBE, vdW-DF, and vdW-DF2 functionals at $P_{DFT}$=200 GPa, 300 GPa, and 400 GPa respectively.  Structures relaxed with the vdW-DF and vdW-DF2 functionals have lower enthalpies than those generated with PBE, as we might have guessed from the relative energetics and from the previous discussion of bond lengths.  Notice that the QMC pressures for all structures differ from their DFT values.  The magnitude of the deviance follows the ordering we discussed above.

\begin{figure}[h]   
    \includegraphics[scale=0.35]{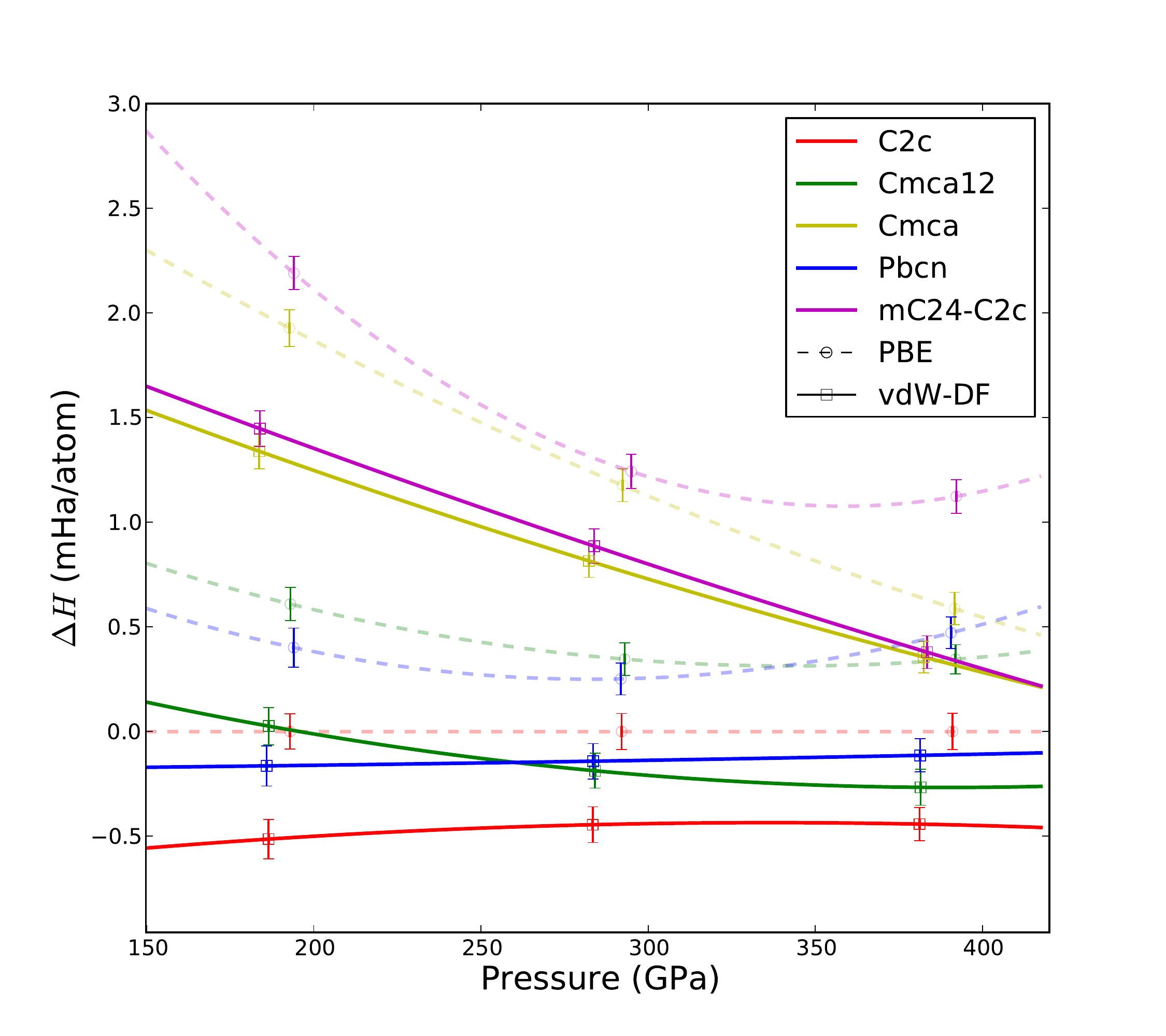}
    \includegraphics[scale=0.35]{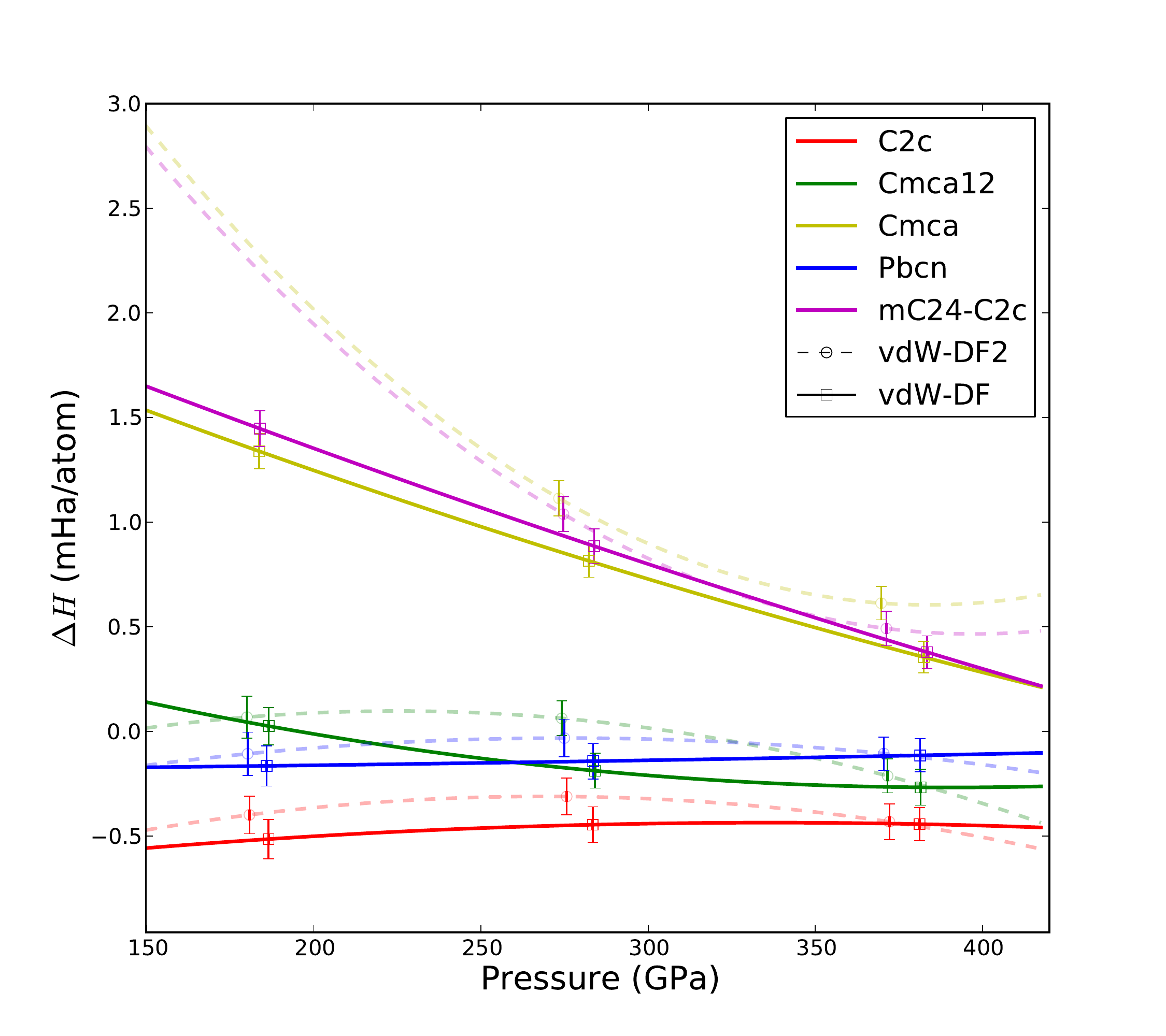}
     \caption{(Color online) (Top) Relative enthalpies of candidate ground-state structures relaxed with either PBE or vdW-DF. Structures relaxed with the vdW-DF functional are shown with bold solid lines, and those with the PBE functional with light dashed lines.  The color of the line denotes the structure.  (Bottom) Comparing vdW-DF and vdW-DF2 functionals.  The $P^{DF}=200GPa$ $Cmca$ and $mC24-C2/c$ structures optimized with the vdW-DF2 functional had significantly lower QMC pressures than the vdW-DF and PBE structures, and so these points are not shown but were used in the fit.  } 
     \label{fig:enthcompare}
\end{figure}

At the top of Figure \ref{fig:enthcompare}, we show the relative enthalpies for structures obtained through PBE and vdW-DF optimization. As in the previous figure, relative deviations from data for C2/c structures optimized with PBE functional are shown. We see that structures optimized using the PBE and vdW-DF functionals exhibit similar qualitative features; the ordering of the ground-state structures is consistent between functionals, as are the pressure trends in the relative enthalpy curves.  However, the vdW-DF structures are all lower in enthalpy than their PBE counterparts, and there are noticeable quantitative differences.  For instance, the relative enthalpy differences between $Cmca$ and $mC24-C2/c$ are much larger in PBE around ~200 GPa and ~400 GPa than in the vdW-DF functional.

At the bottom of Figure \ref{fig:enthcompare}, we compare the relative enthalpy differences between structures optimized with the vdW-DF and vdW-DF2 functionals.  Note that even though the relative enthalpies for vdW-DF2 are all lower than for PBE, they always remain a little higher than for structures optimized using the vdW-DF functional.  Both functionals illustrate the same qualitative trends, although note that the quantitative agreement is much greater than between vdW-DF and PBE. For instance, the relative enthalpy difference between $Cmca$ and $mC24-C2/c$ is much smaller with vdW-DF2 than PBE.  

\subsubsection{Intra-Molecular Potential}

While an accurate bond length is necessary for a precise prediction of the relative electronic enthalpy between different structures, the shape of the intra-molecular potential is equally important for the accurate description of both thermal and quantum ionic components of the enthalpy. From the calculations of the bond length dependence of the energy, we can measure the ability of each DFT functional to reproduce the intra-molecular potential of the hydrogen molecule in various structures. For this purpose, similar to the RQMC calculations presented above, we calculated the bond length dependence of the electronic energy with PBE, vdW-DF and vdW-DF2, on the same structures presented in section \ref{subset:bond_lengths}. From the resulting energies, we obtain the curvature $\alpha$ of the potential at the corresponding equilibrium bond length by fitting a quadratic function. This curvature is directly related to the vibrational frequency of the molecule, which is the leading contribution to the zero-point energy due to its high frequency. Inaccuracies in the curvature of the potential will lead to systematic errors in the zero-point energies calculated with DFT. 

\begin{table*}[t]
\begin{center}
\begin{tabular}{c c c c c c}
\hline
\hline
Pressure ($GPa$) &  $C2/c$   &  $Cmca$   &  $Cmca-12$   &  $Pbcn$    & $mC24-C2/c$   \\
\hline
\hline
 \multicolumn{6}{c}{PBE (24.7)} \\
\hline
200 & -26(3) & -28(5) & -24(5) & -21(4) & -34(5) \\ 
300 & -21(3) & 43(16) & -29(23) & -23(4) & -20(9) \\ 
400 & -25(5) & -12(7) & 19(12) & 3(9) & -42(6) \\ 
%  MAE 20.2 
\hline  
\hline
 \multicolumn{6}{c}{vdW-DF (7.3)} \\
\hline
  %#vdW-DF
200 & -3(2) & -17(2) & -10(2) & -9(1) & -5(3) \\ 
300 & -8(1) & -6(3) & -8(4) & -9(2) & -5(5) \\ 
400 & 9(5) & 5(9) & 0(7) & -4(6) & 11(7) \\ 
  % # MAE 5.6
\hline
\hline
 \multicolumn{6}{c}{vdW-DF2 (19.2)} \\
\hline  
   %# vdW-DF2
200 & 19(5) & 20(6) & - & 27(3) & 37(6) \\ 
300 & 10(3) & 6(7) & 15(7) & 16(4) & 22(8) \\ 
400 & 12(4) & 10(7) & 15(6) & 25(6) & 37(16) \\ 
%# MAE  8.5
\hline
\hline
\end{tabular}
\end{center}
\caption{Comparison of the curvature of the intra-molecular potential of hydrogen between selected DFT functionals and RQMC. The table shows the percentage of difference between RQMC and DFT, calculated as: $PE (\%) = (\alpha_{DFT}-\alpha_{RQMC})/\alpha_{RQMC}*100$, where $\alpha$ is the second derivative of the total energy as a function of the molecular bond length in the solid at the equilibrium distance. The relative mean absolute error, averaged over all structures and all pressures,  is shown in parenthesis next to each functional name.  A missing calculation is represented with a short dashed line.}
\label{table:curv}
\end{table*}%

\begin{figure}[h]   
    \includegraphics[scale=0.4]{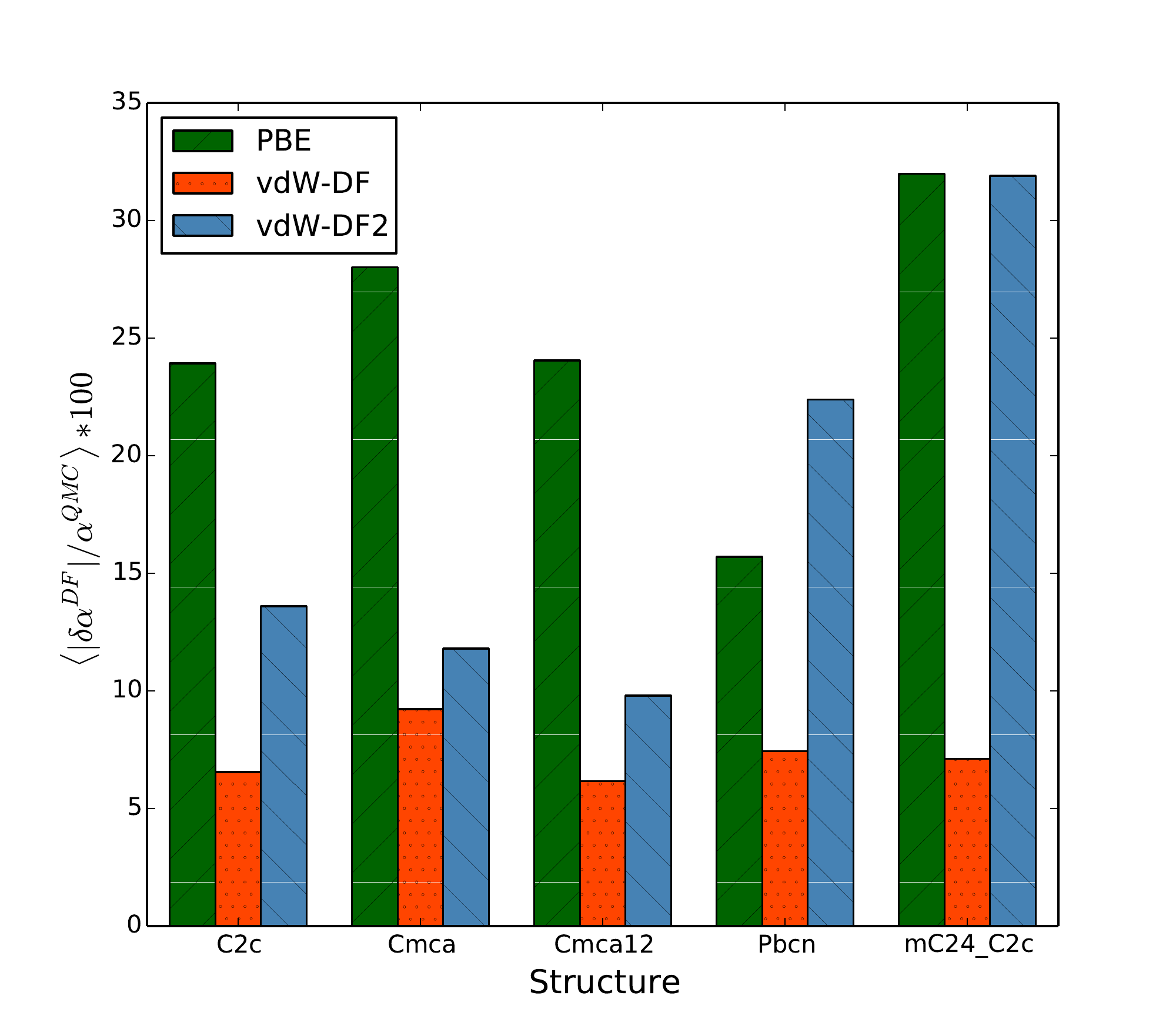}
    \includegraphics[scale=0.4]{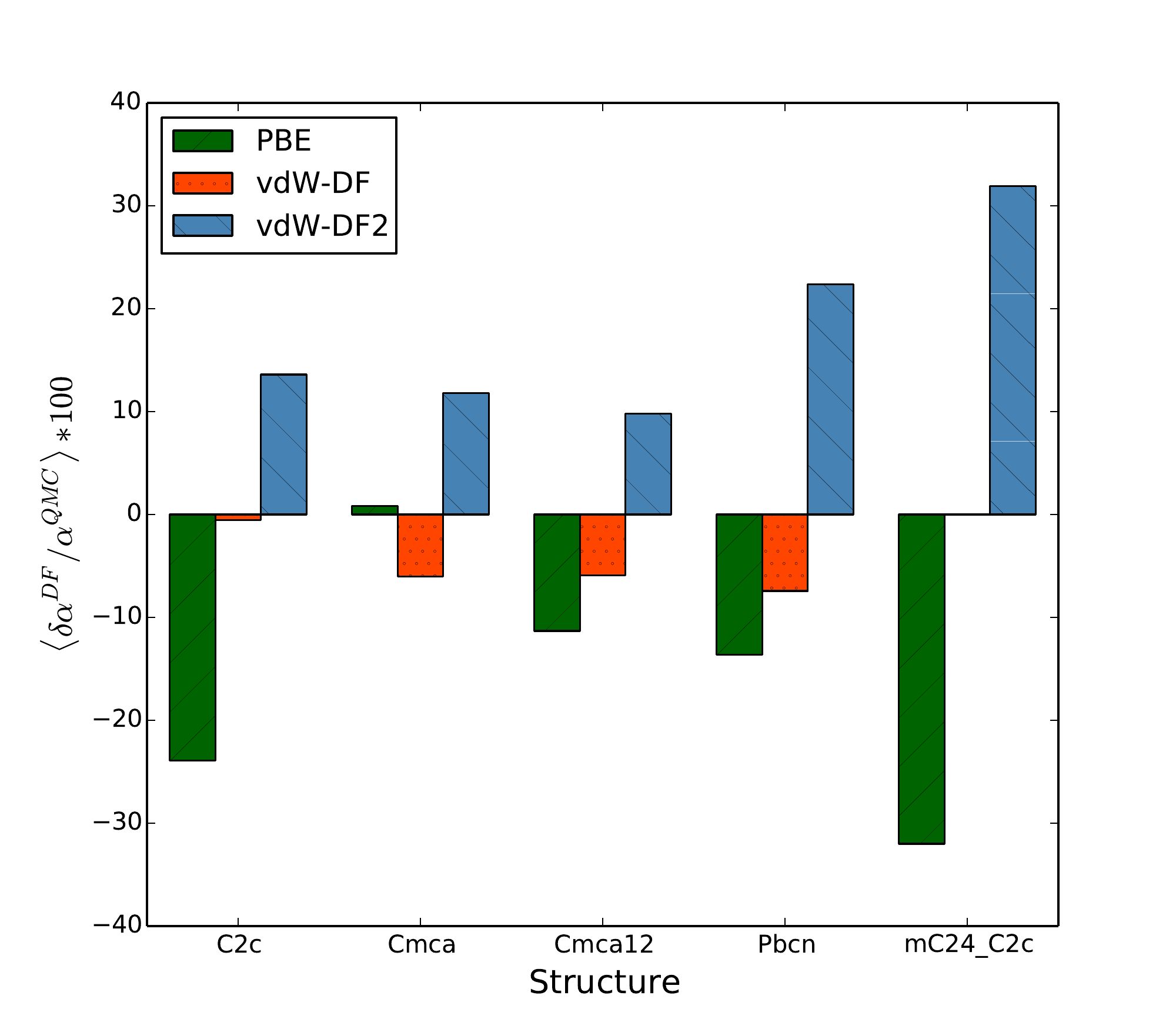}
     \caption{(Color online) Above are two measures of the relative errors in the curvature $\alpha$ for different structures and functionals.  The angle brackets denote averaging over all pressures.  (Top) Relative mean absolute error of the curvature for each functional and structure.  (Bottom) Relative mean error of the curvature. } 
     \label{fig:curvcompare}
\end{figure}

Table \ref{table:curv}  and Figure \ref{fig:curvcompare} show a comparison of the curvatures of the intra-molecular potential between various DFT functionals and RQMC calculations. It is clear that PBE systematically underestimates the magnitude of the curvature, by an average of $\sim$20$\%$ over the studied pressure range. vdW-DF2, on the other hand, systematically overestimates the curvature but by a smaller amount. In both cases, the variation with structure is large. As can be expected, the vdW-DF functional shows a good overall agreement with QMC, producing an average discrepancy of $\sim$5$\%$ over the entire configuration set. This is consistent with the results of section \ref{subsec:local_energetics} that shows that vdW-DF gives a more accurate estimate of the local potential energy surface of the solid and of the relative energies of different molecular configurations, relative to PBE and vdW-DF2. As a consequence, both thermal and zero-point components of the energies should be computed with this functional for a more predictive calculation.  

\subsubsection{Liquid-Liquid Phase Transition (LLPT)}

The location of the LLPT in hydrogen has been recently shown to depend significantly on both NQEs and on the treatment of electronic exchange and correlation \cite{Morales2013}. While the inclusion of NQEs typically reduces the transition pressure by $\sim$60-80 $GPa$, different DFT functionals produce variations by as much as $\sim$200$GPa$. In this section, we use the QMC calculations presented in section \ref{subsec:global_energetics} to show that there is a correlation between the location of the LLPT estimated by a given functional and the difference in the average energy between the QMC energy and a DF energy.

The predicted transition pressure will be affected by differences of errors in the two branches of the free energy $F(T,V)$ isotherm at the transition point. To estimate the principal effect of DF errors, we look at the difference in DF internal energy errors between two different densities corresponding to the atomic liquid and molecular liquid. We begin, as in the global and energetic sections of this paper, by defining a test set $S'$ to be the aggregate of all liquid test configurations at $r_s=1.30,1.45,1.60$.  Choosing $c^{DF}$ to be the median of this aggregated set, we calculate $\langle \widetilde {\delta e^{DF}(r_s) } \rangle$ for $r_s=1.30$ and $r_s=1.60$.  Then  $(\langle \delta e^{DF}\rangle_{at}- \langle \delta e^{DF}\rangle_{mol})$ measures the mean energy shift between the atomic and molecular states. 

We then estimate the transition pressure using several functionals; the procedure for calculating the transition pressure is given in the Supplemental Information.  In Section \ref{subsec:pressures}, we found that there is a systematic and sometimes sizable error in DF pressure estimates, which causes an additional bias of the transition pressures.  We correct for this error by fitting the pressure errors to $\delta P^{DF}(r_s)=a_0+a_1 P^{QMC}(r_s) +a_2[P^{QMC}(r_s)]^2$ for each functional, where the coefficients $a_i$ are assumed to be independent of density.  We then solve this equation for $P^{QMC}$ as a function of $P^{DF}$, which gives us a corrected transition pressure.

\begin{figure}[h]   
    \includegraphics[scale=0.4]{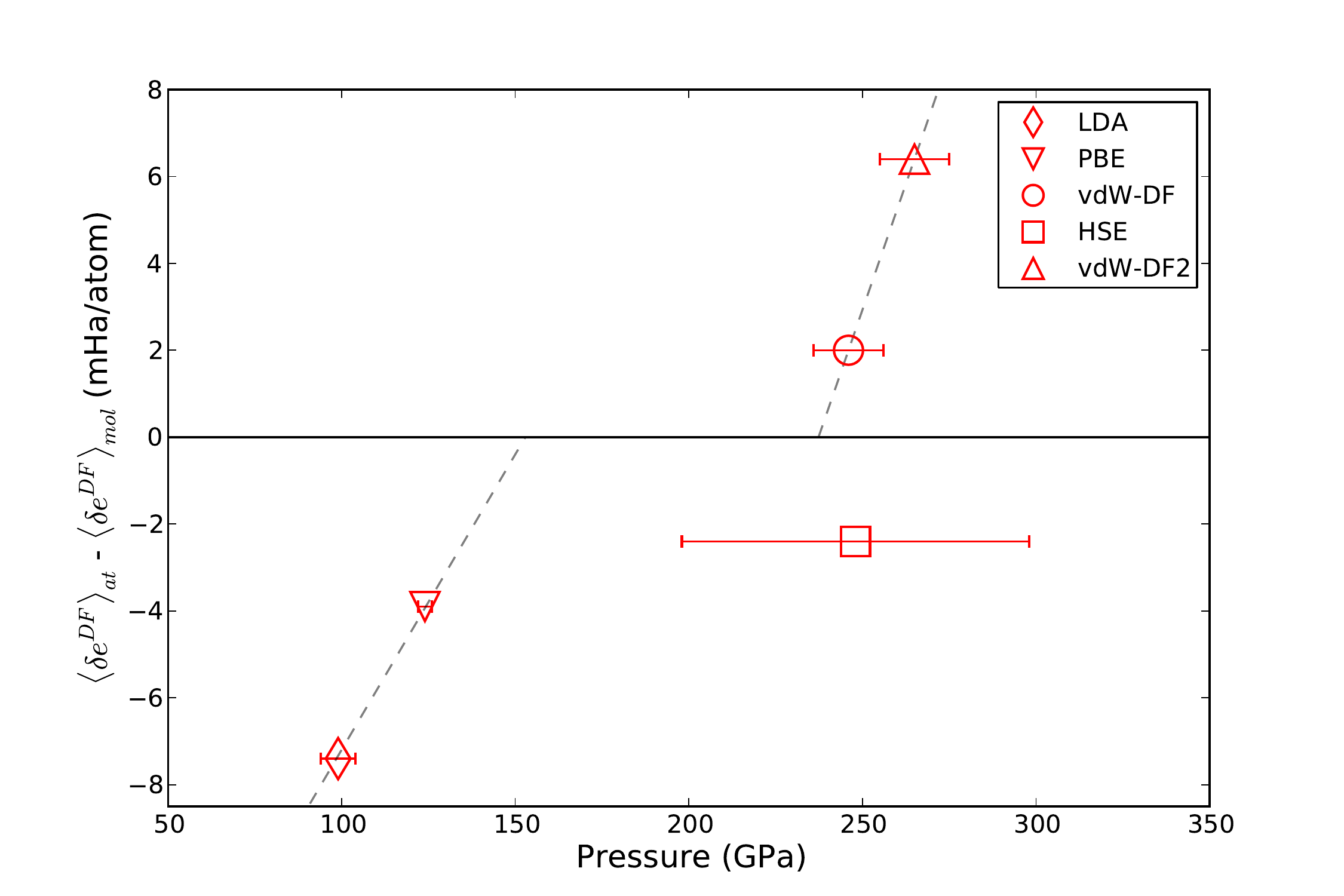}
     \caption{(Color online) $\langle \delta e^{DF}\rangle_{at}$ - $\langle \delta e^{DF}\rangle_{mol}$ for a selection of DFT functionals as a function of estimated $P^{LLPT}$ for 54 protons at T = 1000 K.  The different marker styles indicate different functionals. Dashed lines are linear fits to the positive and negative data points.} 
     \label{fig:llpt_vs_mae}
\end{figure}

In Figure \ref{fig:llpt_vs_mae}, we plot $\langle \delta e^{DF}\rangle_{at}$ - $\langle \delta e^{DF}\rangle_{mol}$ versus the corrected transition pressure for all considered functionals.  We see that the errors change sign as we go from PBE, LDA, and HSE to the van der Waals functionals.  If we knew the functional relationship between the energy errors and the transition pressure, the point where that function crosses the x-axis should coincide with the correct transition pressure for a 54 atom system at 1000K.  Though our data is too sparse to characterize this function, we can attempt to bound the transition.  By performing linear fits on the positive and negative data points independently, we show in Figure \ref{fig:llpt_vs_mae} that the transition probably lies between approximately 150 and 240 GPa.  This is about a 40\% reduction in the transition pressure uncertainty due to electronic correlation pointed out in previous work \cite{Morales2013}.  

We have made preliminary estimates of the transition pressure using the coupled electron-ion Monte Carlo\cite{LNP2006}, a method which treats the electrons with QMC.  Using quantum protons, we estimate the transition pressure for a 54 atom system at T=1000K to be around 221 GPa; details will be published in a future work.  It is reassuring to see that this estimate lies within the transition pressure error bounds obtained in this section.

%%%%%%%%%%%%%%%%%%%%%%%%%%%%%%%%%%%%%%%%%%%%%%%%%%
%\section{Discussion}
%\label{sec:discussion}
%%%%%%%%%%%%%%%%%%%%%%%%%%%%%%%%%%%%%%%%%%%%%%%%%%%
%
% some discussion here.

%%%%%%%%%%%%%%%%%%%%%%%%%%%%%%%%%%%%%%%%%%%%%%%%%%
\section{Summary and Conclusions}
\label{sec:concl}

In this article we presented a detailed benchmark of DFT exchange-correlation functionals in high pressure hydrogen using accurate QMC calculations. Particular care was taken to control systematic errors in the QMC calculations, including size effects, time step, twist averaging, projection times and population control. We find that the performance of most DFT functionals depend on the property being studied. While LDA and HSE consistently produce the best pressures in both solid and liquid phases, vdW-DF is clearly superior in terms of local energy differences in the potential energy surface. HSE and vdW-DF perform equally well in terms of energetics in liquid hydrogen close to metallization, but with errors in opposite directions. In general, PBE does a rather poor job at describing the relative energies of configurations and describes quite poorly the properties of the molecular bond. This leads to a large underestimate of the metallization transition and of both thermal and zero-point energy contributions. Predictions made with this functional are much less accurate. While no functional considered was capable of accurately describing relative energy differences over a large region of the phase diagram, vdW-DF was found to produce excellent results within a given phase, particularly for the reproduction of the intra-molecular potential and equilibrium bond length. This functional should be used to estimate zero-point energy, which is dominated by energy differences close to an equilibrium configuration. 

%%%%%%%%%%%%%%%%%%%%%%%%%%%%%%%%%%%%%%%%%%%%%%%%%%%

% ******************************
% ACKNOWLEDGMENTS
% ******************************
\begin{acknowledgments}
M.\ A.\ M.\ and J.\ M.\ were supported by the U.S. Department of Energy at the Lawrence Livermore National Laboratory under Contract DE-AC52-07NA27344.  MAM, JM, RC and DMC were supported through the Predictive Theory and Modeling for Materials and Chemical Science program by the Basic Energy Science (BES), DOE. R.\ C.\, J.\ M.\ M.\ and D.\ M.\ C.\ were also supported by DOE DE-NA0001789. C.\ P.\ was supported by the Italian Institute of Technology (IIT) under the SEED project Grant 259 SIMBEDD.
Computer time was provided by the US DOE INCITE program, Lawrence Livermore National Laboratory through the 7th Institutional Unclassified Computing Grand Challenge program and PRACE project n 2011050781.
\end{acknowledgments}

\bibliography{benchmark} 

\end{document}

% ******************************
% Title of paper
% ******************************
\title{ Supplementary Information}
\section{Quantum Monte Carlo}
\subsection{Trial Wavefunction}
  We used a Slater-Jastrow type trial wavefunction for all QMC simulations.  All systems were assumed to be spin unpolarized.  We generated the single-particle orbitals in Quantum Espresso,  using a norm-conserving pseudopotential with a cutoff radius of 0.5 bohr.  We generated orbitals for a 6x6x6 Monhkorst-Pack k-point grid with offset, which was found to converge the diffusion Monte Carlo energies with respect to twist-averaging.  A planewave cutoff of 180 Ry was chosen.

The Jastrow factor contained radially-symmetric one body $e-p$ and two-body $e-e$ terms, represented using a fully-optimizable cubic b-splines basis in real space.  All Jastrow factors had real-space cutoff radii, beyond which the function was zero.  The one-body $e-p$ term had two jastrow terms with cutoff radii of L/2 and 1 bohr, each represented with 8 knots.  We were able to satisfy the $e-p$ cusp condition by fixing the cusp of the short-range term.    The two-body $e-e$ consisted of a term for like spins and a term for opposite spins.  Both used 10 knots and a cutoff radius of L/2.  The proper $e-e$ cusp condition was applied to both terms.    

The backflow transformation was also represented using a fully-optimizable b-spline basis.  We included an $e-p$ term, an $e-e$ term for like spins, and an $e-e$ term for unlike spins.  All terms used 8 knots and a cutoff radius of L/2.  The inclusion of the backflow transformation noticeably improved wave-function quality in the solid systems considered.  In the ground-state $C2/c$ structure optimized at $P=200 GPa$ using the PBE functional, the inclusion of backflow improved the energy by 1 $mHa/atom$, reduced the variance from 2.2 $mHa^2/atom$ to 1.4 $mHa^2/atom$, and increased the QMC pressure estimate by about $0.88 GPa$.  These improvements were mostly consistent between structures at the same pressure, but became less significant as pressure increased.  %%%Have a section going into more detail about the improvements associated with backflow?  A table perhaps with some representative configurations?  Or just leave this to supplemental information?

The trial wavefunctions were optimized by energy-minimization with the linear method \cite{Umrigar2007}.  This was accomplished in two stages.  First, we chose a thermal $C2/c$ structure and fully optimized a trial-wavefunction for a single k-point without backflow.  Then we introduced backflow and optimized the backflow transformations and Jastrows simultaneously for that same k-point until convergence was reached.  Once we had this starting point, we performed three energy minimization steps on $60,000$ samples per step for all considered configurations.  We confirmed that all trial wavefunctions had converged to within this study's error bars at the VMC level.  To save time, we did not reoptimize trial wavefunctions for different twists; we found that the wavefunctions showed no statistically significant in the energy variance by doing so.  

\subsection{Systematic Errors}
Concerning time step errors in our DMC simulations, for the solids we used a time step of 0.025 $Ha^{-1}$ in the solids, which gave a 99.2\% acceptance rate.  We assume that the time step errors between different solid configurations at the same pressure should have similar time step errors.  As representative cases, time step errors were quantified by looking at the PBE generated $C2/c$ ground state structures at 200 $GPa$, 300 $GPa$, and 400 $GPa$.  For the DMC energies, the only statistically significant correction was +0.047 $mHa/atom$ at 300 $GPa$.  No other pressures had statistically significant time step errors in the energy.  For the QMC pressure estimates, the only statistically significant correction was found to be -0.41 $GPa$ at 300 GPa, with P=200 and 300 GPa having no statistically significant corrections.  We added these corrections to all solid configurations at the appropriate pressures.

For the liquid configurations, we used a time step of 0.05 $Ha^{-1}$ with an acceptance ratio of 98.2 \%.  We performed a time-step extrapolation on one PBE configuration from each density, and found that the only statistically significant time-step correction to the energy and pressure was for $r_s$=1.30.  We found this correction to be 0.047 $mHa/atom$ for the energy, and -0.27 $GPa$ for the pressure. We used 1024 walkers for both the solid and liquid configurations.  We did not find a statistically significant population bias.  

All correlated sampling calculations with reptation Monte Carlo were performed with a time step of 0.01 $Ha^{-1}$ and 100 time slices in the discretization of the reptile, leading to a projection time of 1.00 Ha. This was found to be enough to converge energy differences as the bond length of the molecules was varied in the different solid configurations. We used twist averaging with a 4x4x4 MP k-point grid. No size corrections were applied to the energy differences, since they were found to cancel almost exactly.

\subsection{Finite Size Errors}
To control for finite-size errors in the QMC calculations, we used twist-averaging and 1/N finite-size extrapolations.  For twist-averaging, we used a 3x3x3 Monkhorst-Pack (MP) k-point grid for the liquid configurations, and a 6x6x6 grid for the solids.

For both liquid and solid systems, a $1/N$ extrapolation was also performed on both the DMC energies and pressure estimates.  For each ground-state structure, we generated 1x1x1 and 2x2x2 supercells.  For both supercells, we used the optimized trial wavefunctions that neglected the backflow transformation.  The finite-size extrapolations were then performed on the DMC energies and extrapolated estimates of the QMC pressures.  We believe this correction to be reasonable for these disordered structures because for DFT and QMC, we are studying of 54 or 96 atom systems in periodic boundary conditions, and not true liquids and solids in the thermodynamic limit.  

To handle the solid thermal configurations, we took the finite size corrections to be the same as for the corresponding ground-state structures.  For the liquid configurations, we took 5 independent configurations for each density, obtained the finite size corrections for each with a 1/N extrapolation, and averaged the result to obtain the finite-size corrections for each density.

We investigated the use of the Kwee, Zhang, and Krakauer (KZK) DFT based finite-size correction correction scheme for the solid hydrogen systems \cite{Kwee2008, Azadi2013a}, but found that the energy difference between the KZK correction and our 1/N extrapolation could be as much as $\pm 0.5 mHa/atom$ when looking at relative energies between different structures.  This amounts to at most only about $10\%$ of the total finite-size correction, but is significant given that the enthalpy difference between proposed ground-state structures are comparable to this error.

\section{Density Functional Theory}

Calculations using LDA, PBE\cite{Perdew1981,Perdew1996}, vdW-DF \cite{Dion2004}, and vdW-DF2 \cite{Lee2010} functionals were done in Quantum Espresso\cite{Giannozzi2009} with a plane-wave cutoff of 200 Ry and a hard norm-conserving pseudopotential with a cutoff radius of 0.5 Bohr.  Calculations using vdW-optPBE, vdW-optB88 \cite{Klimes2010}, vdW-optB86B \cite{Klimes2011}, BLYP \cite{Grimme2006}, vdW-TS \cite{Tkatchenko2009}, and HSE  \cite{Heyd2003} functionals were done in a modified version of VASP 5.3.3 \cite{Kresse1993, Kresse1996, Kresse1999} with a planewave cutoff of 58.8 Ry and a projector augmented wave formulation \cite{Blochl1994} with a real-space cutoff of 1.10 bohr.  All calculations were performed on a 6x6x6 Monkhorst-Pack grid with offset.

\section{Results}
\subsection{Global and Local Energetics}
In Tables \ref{Table:solid_data} and \ref{Table:liquid_data}, we show the raw global and local energetic errors over all test sets mentioned in the main text.  

\begin{table}[t]{Solid Test Set Error Measurements}
\begin{center}
\begin{tabular}{c c c c c c c c c c c c}
\hline
\hline
Structure &  $P^{DF}$(GPa)  &  LDA   &  PBE & vdW-DF & vdW-DF2 & HSE & vdW-TS & optPBE & optB86B & optB88 & BLYP   \rule{0pt}{0.4cm} \\[0.1cm]
\hline
\hline
\multicolumn{12}{c}{$\langle | \widetilde{\delta e^{DF}} | \rangle_{global} $ (mHa/proton)}  \rule{0pt}{0.4cm} \\[0.1cm]
\hline
C2/c & 200 & 0.461 & 0.371 & 0.131 & 0.168 & 0.353 & 0.389 & 0.328 & 0.372 & 0.344 & 0.166 \\ 
Cmca & 200 & 0.276 & 0.188 & 0.402 & 0.499 & 0.145 & 0.189 & 0.199 & 0.166 & 0.184 & 0.366 \\ 
Cmca-12 & 200 & 0.659 & 0.342 & 0.059 & 0.251 & 0.086 & 0.378 & 0.075 & 0.249 & 0.076 & 0.072 \\ 
Pbcn & 200 & 0.430 & 0.374 & 0.137 & 0.121 & 0.354 & 0.376 & 0.331 & 0.374 & 0.346 & 0.178 \\ 
\hline
C2/c & 300 & 0.445 & 0.309 & 0.244 & 0.187 & 0.255 & 0.338 & 0.266 & 0.309 & 0.274 & 0.260 \\ 
Cmca & 300 & 0.382 & 0.277 & 0.209 & 0.321 & 0.235 & 0.313 & 0.216 & 0.233 & 0.212 & 0.201 \\ 
Cmca-12 & 300 & 0.451 & 0.302 & 0.057 & 0.167 & 0.165 & 0.314 & 0.136 & 0.236 & 0.131 & 0.063 \\ 
Pbcn & 300 & 0.333 & 0.229 & 0.187 & 0.195 & 0.150 & 0.248 & 0.183 & 0.211 & 0.190 & 0.201 \\ 
\hline  
\hline
\multicolumn{12}{c}{$\langle | \widetilde{\delta e^{DF}} | \rangle_{local}$  (mHa/proton)}  \rule{0pt}{0.4cm} \\[0.1cm]
\hline
C2/c & 200 & 0.320 & 0.173 & 0.032 & 0.166 & 0.089 & 0.184 & 0.049 & 0.137 & 0.054 & 0.022 \\ 
Cmca & 200 & 0.275 & 0.151 & 0.042 & 0.155 & 0.078 & 0.160 & 0.049 & 0.120 & 0.052 & 0.034 \\ 
Cmca-12 & 200 & 0.234 & 0.140 & 0.047 & 0.111 & 0.070 & 0.147 & 0.063 & 0.114 & 0.065 & 0.054 \\ 
Pbcn & 200 & 0.262 & 0.151 & 0.036 & 0.121 & 0.085 & 0.157 & 0.069 & 0.123 & 0.072 & 0.051 \\ 
\hline
C2/c & 300 & 0.314 & 0.182 & 0.037 & 0.185 & 0.091 & 0.196 & 0.052 & 0.137 & 0.051 & 0.030 \\ 
Cmca & 300 & 0.360 & 0.222 & 0.032 & 0.174 & 0.105 & 0.239 & 0.085 & 0.175 & 0.083 & 0.037 \\ 
Cmca-12 & 300 & 0.343 & 0.213 & 0.041 & 0.166 & 0.101 & 0.230 & 0.080 & 0.168 & 0.077 & 0.042 \\ 
Pbcn & 300 & 0.306 & 0.187 & 0.055 & 0.166 & 0.089 & 0.200 & 0.074 & 0.146 & 0.073 & 0.054 \\ 
\hline  
\hline
\multicolumn{12}{c}{$\langle \delta P^{DF}\rangle$ (GPa)} \rule{0pt}{0.4cm} \\[0.1cm]
\hline
C2/c & 200 & -0.758 & 3.314 & 10.857 & 17.322 & 1.097 & -8.794 & 0.246 & 4.079 & 9.169 & 10.086 \\ 
Cmca & 200 & 1.823 & 5.960 & 13.474 & 19.852 & 3.772 & -6.244 & 0.335 & 6.678 & 11.769 & 12.647 \\ 
Cmca-12 & 200 & 0.574 & 4.451 & 12.140 & 19.102 & 2.105 & -7.528 & 0.291 & 5.375 & 10.551 & 11.574 \\ 
Pbcn & 200 & -1.020 & 3.070 & 10.662 & 17.179 & -4.411 & -9.006 & 0.239 & 3.831 & 8.952 & 9.872 \\ 
\hline
C2/c & 300 & -0.189 & 3.867 & 12.992 & 22.299 & 1.263 & -12.991 & 0.297 & 5.044 & 11.930 & 12.962 \\ 
Cmca & 300 & 3.305 & 7.387 & 16.510 & 25.819 & 4.821 & -9.691 & 0.420 & 8.655 & 15.647 & 16.613 \\ 
Cmca-12 & 300 & 1.920 & 5.827 & 14.950 & 24.447 & 3.080 & -11.080 & 0.365 & 7.069 & 14.004 & 15.063 \\ 
Pbcn & 300 & -0.400 & 3.621 & 12.776 & 22.147 & -5.358 & -13.284 & 0.289 & 4.819 & 11.729 & 12.768 \\
\hline
\hline
\end{tabular}
\end{center}

\caption{Measures of $\langle | \widetilde{\delta e^{DF}} | \rangle_{global}$, $\langle | \widetilde{\delta e^{DF}} | \rangle_{local} $, and $\langle \delta P^{DF}\rangle$ for all solid test sets.  The test set is identified by the first two columns, which denote the solid structure and DFT pressure respectively.  The subsequent columns denote the error associated with various density functionals.  The error measure and units are given by appropriate headers. }
\label{Table:solid_data}
\end{table}%

\setlength{\tabcolsep}{0.35cm}
\begin{table}[t]{Liquid Test Set Error Measurements}
\begin{center}
\begin{tabular}{c c c c c c}
\hline
\hline
$r_s$  &      LDA      &     PBE     & vdW-DF & vdW-DF2 & HSE   \\
\hline
\hline
 \multicolumn{6}{c}{$\langle | \widetilde{\delta e^{DF}} | \rangle_{global}$  (mHa/proton)} \rule{0pt}{0.4cm} \\[0.1cm]
\hline
1.30 & 0.310 & 0.207 & 0.118 & 0.250 & 0.141 \\ 
1.45 & 0.397 & 0.228 & 0.088 & 0.207 & 0.113 \\ 
1.60 & 0.371 & 0.178 & 0.072 & 0.181 & 0.099 \\
\hline  
\hline
 \multicolumn{6}{c}{$\langle \delta P^{DF} \rangle$  (GPa)} \rule{0pt}{0.4cm} \\[0.1cm]
 \hline
1.30 & 6.386 & 9.613 & 19.891 & 31.108 & -0.572 \\ 
1.45 & 4.168 & 8.304 & 15.907 & 22.435 & 1.597 \\ 
1.60 & 0.910 & 4.388 & 9.351 & 11.820 & -0.735 \\ 
\hline
\hline

\end{tabular}
\end{center}

\caption{Measures of  $\langle | \widetilde{\delta e^{DF}} | \rangle_{global}$ and $\langle \delta P^{DF}\rangle$ for the liquid test sets.  Each test set is identified by an $r_s$ density. }
\label{Table:liquid_data}
\end{table}%

\subsection{Estimating the LLPT}
The transition pressures for all functionals except HSE were calculated using PIMD simulations for a system of 54 atoms with a shifted 2x2x2 MP k-point grid, along the T=1000K isotherm.  PIMD simulations were run on a grid of densities, and the data used to construct a $P-\rho$ isotherm.  The transition pressure was estimated as the value of the observed plateau, as in previous works \cite{MoralesPNAS2010}. The uncertainty in this estimate results from the use of a finite grid of densities in the calculation of the isotherm. 

In the case of the simulations with HSE, molecular dynamics simulations with classical protons were performed at either the $\Gamma$-point  of the simulation cell.  This was done due to the much higher time requirements of HSE calculations. The corresponding transition pressure was then shifted by 60 $GPa$ in order to account for NQEs.  We observe a strong variation on the location of the LLPT with the choice of k-points when we use HSE. This is not unexpected, since the exchange potential depends on the grid of k-points used in the calculation and leads to very slow convergence. As a consequence, the uncertainty on the location of the HSE transition is quite large.

\begin{table}[t]{Transition Pressures vs. Global Energy Errors}
\begin{center}
\begin{tabular}{c | c| c }
\hline
\hline
Functional &  $P_{LLPT} (GPa)$ &  $\langle \delta e^{DF}\rangle_{at}$ - $\langle \delta e^{DF}\rangle_{mol}$ (mHa/proton)  \rule{0pt}{0.4cm} \\[0.1cm]
\hline
\hline
LDA &  99(5) & -7.4  \\ 
PBE &  124(2) & -3.9 \\
vdW-DF  & 246(10) & 2.0 \\
vdW-DF2  & 265(10) & 6.4 \\
HSE & 248(50) & -2.4 \\
\end{tabular}
\end{center}
\caption{Table of $\langle \delta e^{DF}\rangle_{at}$ - $\langle \delta e^{DF}\rangle_{mol}$ vs estimated transition pressure for all considered functionals. }
\label{Table:llpt_fit}

\end{table}
Table \ref{Table:llpt_fit} shows the raw data for the predicted transition pressure $P_{LLPT}$ and the energy difference registered on the liquid sets as explained in the main text..  To obtain the two linear extrapolations to zero shown in Figure 10 of the main text, we divide the data into two sets based on whether the energy error difference is positive or negative.  A linear fit is then performed on each of these data sets separately.

\bibliography{benchmark}